\documentclass[12pt,a4paper,twoside,fleqn]{article}
\usepackage{latexsym}
\usepackage{epsf}
\usepackage{graphicx}

\usepackage{color}
\newcommand{\g}[1]{\color{red} #1}

\def\be{\begin{equation}}
\def\ee{\end{equation}}
\def\bea{\begin{eqnarray}}
\def\eea{\end{eqnarray}}
\def\nnb{\nonumber}
\def\bbuildrel#1_#2^#3{\mathrel{\mathop{\kern 0pt#1}\limits_{#2}^{#3}}}
\def\slash#1{\setbox0=\hbox{$#1$}#1\hskip-\wd0\dimen0=5pt\advance
       \dimen0 by-\ht0\advance\dimen0 by\dp0\lower0.5\dimen0\hbox
         to\wd0{\hss\sl/\/\hss}}

\newcommand{\scs}{\scriptscriptstyle}
\newcommand{\lo}{\ln 2}
\newcommand{\f}{\frac}
\newcommand{\al}{\alpha_s}
\newcommand{\e}{\epsilon}
\newcommand{\me}[1]{\langle#1\rangle}
\newcommand{\TT}{\rule[-2mm]{0mm}{7mm}}

\newcommand{\mc}[3]{\multicolumn{#1}{#2}{#3}}

\def\theequation{\thesection.\arabic{equation}}
\newcommand{\newsection}[1]{\section{#1}\setcounter{equation}{0}}

\begin{document}
\begin{titlepage}

\begin{flushright}
  {\bf TUM-HEP-371/00\\       
       CERN-TH/2000-124\\
       IFT-13/2000\\
       hep-ph/0005183
%\\  version: \today
}\\[1cm]
\end{flushright}

\begin{center}

\setlength {\baselineskip}{0.3in} 
{\bf\Large Two-Loop QCD Anomalous Dimensions\\ 
of Flavour-Changing Four-Quark Operators\\
  Within and Beyond the Standard Model}\\[2cm]

\setlength {\baselineskip}{0.2in}
{\large  Andrzej J. Buras$^{^{1}}$, 
         Miko{\l}aj Misiak$^{^{2,\,3}}$
         and J{\"o}rg Urban$^{^{1}}$}\\[5mm]

$^{^{1}}${\it Physik Department, Technische Universit{\"a}t M{\"u}nchen,\\
              D-85748 Garching, Germany}\\[3mm]

$^{^{2}}${\it Theory Division, CERN, CH-1211 Geneva 23, Switzerland}\\[3mm]

$^{^{3}}${\it Institute of Theoretical Physics, Warsaw University,\\
                 Ho\.za 69, PL-00-681 Warsaw, Poland}\\[2cm]

{\bf Abstract}\\
\end{center} 
\setlength{\baselineskip}{0.2in} 

We calculate the two-loop QCD anomalous dimension matrix (ADM)
$(\hat\gamma^{(1)})_{\rm NDR}$ in the NDR--$\overline{\rm MS}$ scheme
for all the flavour-changing four-quark dimension-six operators that
are relevant in both the Standard Model and its extensions.  Both 
current--current and penguin diagrams are included. Some of
our NDR--$\overline{\rm MS}$ results for $\Delta F=1$ operators overlap
with the previous calculations, but several others have never been
published before.  In the case of $\Delta F=2$ operators, our results
are compatible with the ones obtained by Ciuchini et al. in the
Regularization-Independent renormalization scheme, but differ from
their NDR--$\overline{\rm MS}$ results. In order to explain the
difference, we calculate the ADM of $\Delta F=2$ operators again,
extracting it from the ADM of $\Delta F=1$ operators.

\end{titlepage} 

\setlength{\baselineskip}{0.3in}

\newsection{Introduction}
\label{sec:intro}

Renormalization group short-distance QCD effects play an important
role in the phenomenology of non-leptonic weak transitions of $K$-,
$D$- and $B$-mesons.  An essential ingredient in any renormalization
group analysis is the anomalous dimension matrix (ADM), which
describes the mixing of the relevant local four-quark operators under
renormalization \cite{BBL96,B98}.

The operators considered in the present paper have the form
\be \label{generic.four.quark} 
\bar \Psi_1^\alpha\,\Gamma_A^k\,\Psi_2^\alpha\;
\bar \Psi_3^\beta\,\Gamma^k_B\,\Psi_4^\beta,\qquad
\bar \Psi_1^\alpha\,\Gamma_A^k\,\Psi_2^\beta\;
\bar \Psi_3^\beta\,\Gamma^k_B\,\Psi_4^\alpha,
\ee
where $\alpha$, $\beta$ are colour indices and $\Gamma_{A,B}^k$ are
generic Dirac matrices given explicitly below. The subscripts $i$ in
$\Psi_i$ are flavour indices. In the case of FCNC transitions with
$\Delta F=2$, such as neutral meson mixing, one has
\be
\Psi_1 = \Psi_3,\qquad\Psi_2 = \Psi_4.
\label{spinorconstr}
\ee
Known examples are the operators $(\bar{s}d)_{V-A}(\bar{s}d)_{V-A}$
and $(\bar{b}d)_{V-A}(\bar{b}d)_{V-A}$ relevant in the Standard Model
(SM) to $K^0$--$\bar{K}^0$ and $B_d^0$--$\bar{B}_d^0$
mixing, respectively.

Four-quark operators that occur in the SM calculations of
flavour-changing processes do not form a complete set of all the
dimension-six four-quark operators.  Other operators need to be
considered in many extensions of the SM, e.g. in the Supersymmetric
Standard Model (SSM) (see e.g.  ref. \cite{GGMS96}). For instance, the
SSM and SM predictions for $K^0$--$\bar{K}^0$ and
$B_d^0$--$\bar{B}_d^0$ mixing can have similar precision only if the
two-loop ADM for {\em all} the $\Delta F=2$ operators is known.

The main purpose of the present paper is a calculation of the two-loop
ADM for all the dimension-six flavour-changing four-quark operators in
the NDR--$\overline{\rm MS}$ scheme ($\overline{\rm MS}$ scheme with
fully anticommuting $\gamma_5$). Our main findings are the
NDR--$\overline{\rm MS}$ anomalous dimensions of the operators with
Dirac structures (cf. eq.~(\ref{generic.four.quark})):
\be \label{newstr}
\Gamma_A^k \otimes \Gamma^k_B = (1\pm\gamma_5) \otimes (1\pm\gamma_5) 
\hspace{1cm} {\rm and} \hspace{1cm} 
\Gamma_A^k \otimes \Gamma^k_B = 
[\sigma_{\mu\nu} (1\pm\gamma_5)] \otimes [\sigma^{\mu\nu} (1\pm\gamma_5)].~~~
\ee
For these operators, our two-loop results differ from the
NDR--$\overline{\rm MS}$ ones of Ciuchini et al.~\cite{CFLMSS97}, but
are compatible with their RI-scheme ADM. For all the other operators,
no new calculation is actually necessary --- all the two-loop results
can be extracted from the existing Standard Model ones.  

Our paper is organized as follows. In section~\ref{sec:df2}, we
perform a direct calculation of the NDR--$\overline{\rm MS}$-scheme ADM
of $\Delta F=2$ operators. This is a relatively straightforward
computation, since all the methods are already known from similar SM
calculations (see e.g.  refs.~\cite{BW90}--\cite{CMM98.df}).
The only novelty here is the introduction of evanescent operators that
vanish by the Fierz identities.

In section~\ref{sec:df1}, we compute the NDR--$\overline{\rm MS}$ ADM
for such $\Delta F=1$ operators, to which only the current--current
diagrams are relevant. Some of the $\Delta F=1$ results have never
been published before.  The ones that are not new agree with the old SM
calculations.  The subject of section~\ref{sec:penguin} are $\Delta
F=1$ operators containing one quark--antiquark pair of the same
flavour. We identify the operators to which the so-called penguin
diagrams are relevant, and give the corresponding anomalous
dimensions.

In section~\ref{sec:RI}, we derive the matrix $\Delta \hat{r}$
necessary for transforming the Wilson coefficients from the
NDR--$\overline{\rm MS}$ to the RI scheme (originally called the MOM
scheme) that is more useful for non-perturbative calculations of
hadronic matrix elements \cite{BMRGGLM00}.

Section~\ref{sec:transf} is devoted to performing a consistency check
of our $\Delta F=1$~ and $\Delta F=2$~ results.  The current--current
ADM of $\Delta F=1$~ operators is transformed there to such an
operator basis, in which the $\Delta F=2$ results can be easily read
off. This calculation serves also as a preparation for the comparison
with Ciuchini et al.~\cite{CFLMSS97}. Comparison with this article and
other existing literature is the subject of section~\ref{sec:comp}. We
conclude in section \ref{sec:concl}.

In appendix A, we list the evanescent operators relevant to the
$\Delta F=2$ calculation. In appendix B, an analogous list for the
$\Delta F=1$ case is presented.  Appendix C contains additional
evanescent operators that become important {\em only} when one wants
to derive the $\Delta F=2$ results from the $\Delta F=1$ ones, as in
section \ref{sec:transf}.  Appendix D is devoted to recalling and
generalizing the notion of ``Greek projections''. Appendix E
contains a list of separate contributions from different diagrams to the one-
and two-loop ADMs for $\Delta F=1$ operators with Dirac structures
(\ref{newstr}). Finally, in appendix F, we outline our determination
of two-loop mixing via penguin diagrams that involves beyond-SM operators.

\newpage
\newsection{Direct calculation of the ADM in the $\Delta F=2$ case}
\label{sec:df2}

For definiteness, we shall consider here operators responsible for the
$K^0$--$\bar{K}^0$ mixing. There are 8 such operators of dimension 6.
They can be split into 5 separate sectors, according to the chirality
of the quark fields they contain. The operators belonging to the first
three sectors (VLL, LR and SLL) read
\bea 
Q_1^{\rm VLL} &=& (\bar{s}^{\alpha} \gamma_{\mu}    P_L d^{\alpha})
              (\bar{s}^{ \beta} \gamma^{\mu}    P_L d^{ \beta}),
\nnb\\[4mm] 
Q_1^{\rm LR} &=&  (\bar{s}^{\alpha} \gamma_{\mu}    P_L d^{\alpha})
              (\bar{s}^{ \beta} \gamma^{\mu}    P_R d^{ \beta}),
\nnb\\
Q_2^{\rm LR} &=&  (\bar{s}^{\alpha}                 P_L d^{\alpha})
              (\bar{s}^{ \beta}                 P_R d^{ \beta}),
\nnb\\[4mm]
Q_1^{\rm SLL} &=& (\bar{s}^{\alpha}                 P_L d^{\alpha})
              (\bar{s}^{ \beta}                 P_L d^{ \beta}),
\nnb\\
Q_2^{\rm SLL} &=& (\bar{s}^{\alpha} \sigma_{\mu\nu} P_L d^{\alpha})
              (\bar{s}^{ \beta} \sigma^{\mu\nu} P_L d^{ \beta}),
\label{normal}
\eea
where $\sigma_{\mu\nu} = \f{1}{2} [\gamma_{\mu}, \gamma_{\nu}]$ and
$P_{L,R} =\f{1}{2} (1\mp \gamma_5)$. The operators belonging to the
two remaining sectors (VRR and SRR) are obtained from $Q_1^{\rm VLL}$ and
$Q_i^{\rm SLL}$ by interchanging $P_L$ and $P_R$. Since QCD preserves
chirality, there is no mixing between different sectors.  Moreover,
the ADMs in the VRR and SRR sectors are the same as in the VLL and SLL
sectors, respectively.  In the following, we shall consider only the
VLL, LR and SLL sectors.

In dimensional regularization, the four-quark operators from
eq.~(\ref{normal}) mix at one loop into the evanescent operators
listed in appendix A. Specifying these evanescent operators is
necessary to make precise the definition of the NDR--$\overline{\rm
  MS}$ scheme in the effective theory \cite{BW90,CMM98.df,DG91,HN95}.
An important novelty in the present case (when compared to $\Delta
F=1$ calculations) is the necessity of introducing evanescent
operators that vanish in 4 dimensions by the Fierz identities. The
Fierz identities cannot be analytically continued to $D$ dimensions.
Therefore, they have to be treated in dimensional regularization in
the same manner as the identity
\be \label{three.gammas}
\gamma_{\mu}\gamma_{\nu}\gamma_{\rho} = 
  g_{\mu\nu} \gamma_{\rho}
+ g_{\nu\rho} \gamma_{\mu}
- g_{\mu\rho} \gamma_{\nu}
+ i \epsilon_{\alpha\mu\nu\rho} \gamma^{\alpha} \gamma_5,
\ee
i.e. appropriate evanescent operators have to be introduced.

As an example, consider the operators $Q_1^{\rm SLL}$ and $Q_2^{\rm SLL}$. When
these operators are inserted into one- and two-loop diagrams, the operators
\bea
\tilde Q_1^{\rm SLL} &=& (\bar s^\alpha P_L d^\beta )
                     (\bar s^\beta  P_L d^\alpha),\\
\tilde Q_2^{\rm SLL} &=& (\bar s^\alpha \sigma_{\mu\nu} P_L d^\beta )
                     (\bar s^\beta  \sigma^{\mu\nu} P_L d^\alpha)
\eea
are generated. In 4 dimensions these operators can be expressed through
$Q_1^{\rm SLL}$ and $Q_2^{\rm SLL}$ by using the Fierz identities 
\bea
(P_L)_{ij} (P_L)_{kl} &=& \f{1}{2} (P_L)_{il} (P_L)_{kj} 
-\f{1}{8} (\sigma_{\mu\nu} P_L)_{il} (\sigma^{\mu\nu} P_L)_{kj}, \nnb\\
(\sigma_{\mu\nu} P_L)_{ij} (\sigma^{\mu\nu} P_L)_{kl} 
&=& -6 (P_L)_{il} (P_L)_{kj} 
-\f{1}{2} (\sigma_{\mu\nu} P_L)_{il} (\sigma^{\mu\nu} P_L)_{kj},
\label{fSLL}
\eea
which give
\bea \label{opfSLL1}
\tilde Q_1^{\rm SLL} &\bbuildrel{=}_{D=4}^{}& -\frac12\,Q_1^{\rm SLL}+\frac18\,Q_2^{\rm SLL},\\
\tilde Q_2^{\rm SLL} &\bbuildrel{=}_{D=4}^{}& 6\,Q_1^{\rm SLL}+\frac12\,Q_2^{\rm SLL}. 
\eea
These relations can be used in the calculation of one-loop ADM. In the   
case of two-loop calculations, in the NDR--$\overline{\rm MS}$ scheme, where
Dirac algebra has to be performed in $D\neq4$ dimensions, these relations
have to be generalized to
\bea
\label{evanops1}
\tilde Q_1^{\rm SLL} &=& -\frac12\,Q_1^{\rm SLL}+\frac18\,Q_2^{\rm SLL}+E_1^{\rm SLL},\\
\label{evanops2}
\tilde Q_2^{\rm SLL} &=& 6\,Q_1^{\rm SLL}+\frac12\,Q_2^{\rm SLL}+E_2^{\rm SLL}.
\eea
Here, $E_1^{\rm SLL}$ and $E_2^{\rm SLL}$ are the evanescent operators that
vanish in 4 dimensions by Fierz identities. They are simply defined by
(\ref{evanops1}) and (\ref{evanops2}) and are given in appendix A.

The effective Lagrangian can be written separately for each sector. 
It takes the form
\be \label{Leff}
{\cal L}_{eff} = -\f{G_F^2 M_W^2}{4 \pi^2} (V^*_{ts} V_{td})^2 
Z_q^2 \sum_i C_i(\mu)
\left[Q_i \;\; + \;\; (\mbox{counterterms})_i \right],
\ee
where $Z_q$ is the quark wave-function renormalization constant.

The coefficients $C_i(\mu)$ satisfy the Renormalization Group
Equation (RGE)
\be \label{RGE}
\mu \f{d}{d \mu} \vec{C}(\mu) = \hat{\gamma}(\mu)^T \vec{C}(\mu) 
\ee
governed by the ADM $\hat{\gamma}(\mu)$ that has the following
perturbative expansion:
\be \label{gamma.exp}
\hat{\gamma}(\mu) = \f{\al  (\mu)}{ 4 \pi   } \hat{\gamma}^{(0)}
                  \;+\; \f{\al^2(\mu)}{(4 \pi)^2} \hat{\gamma}^{(1)}
                  \;+\; {\cal O}(\al^3).
\ee

The ADM in the MS or $\overline{\rm MS}$ scheme is found from one-
and two-loop counterterms in the effective theory, according to the
following relations (equivalent to eqs.~(4.26)--(4.37) of
ref.~\cite{BW90}):
\bea
\hat{\gamma}^{(0)} &=& 2 \hat{a}^{11}, \label{gamma0} \\
\hat{\gamma}^{(1)} &=& 4 \hat{a}^{12} - 2 \hat{b} \hat{c}.
\label{gamma1}
\eea
The matrices $\hat{a}^{11}$, $\hat{a}^{12}$ and $\hat{b}$ in the above
equations parametrize the MS-scheme counterterms in
eq.~(\ref{Leff}) (for $D=4-2\e$)
\bea \label{counterterms}
(\mbox{counterterms})_i &=& \vspace{0.2cm}
\f{\al}{4 \pi \e} \left[ \sum_k a^{11}_{ik} Q_k 
                   \;+\; \sum_k b_{ik} E_k \right] 
\;+\; \f{\al^2}{(4 \pi)^2} \sum_k
\left( \f{1}{\e^2} a^{22}_{ik} + \f{1}{\e} a^{12}_{ik} \right) Q_k 
\nnb \\[2mm] 
&+& \; (\mbox{two-loop evanescent counterterms}) + {\cal O}(\al^3). 
\eea

The matrix $\hat{c}$ is recovered from one-loop matrix elements of the
evanescent operators. Let us denote by $\me{E_k}_{\rm 1loop}$ the one-loop
$K^0$--$\bar{K^0}$ amplitude with an insertion of some evanescent
operator $E_k$. The pole part of such an amplitude is proportional to
some linear combination of tree-level matrix elements of evanescent
operators. The remaining part in the limit $D \to 4$ can be expressed
by tree-level matrix elements of the physical operators $Q_i$. The
finite coefficients of these matrix elements define the matrix
$\hat{c}$ as follows:
\bea \label{evan.melem}
\me{E_k}_{\rm 1loop} &=& \;-\; \f{1}{\e} \left[ 
\sum_j d_{kj} \me{E_j}_{\rm tree} + 
\sum_j e_{kj} \me{F_j}_{\rm tree} \right]
\;-\; \sum_i c_{ki} \me{Q_i}_{\rm tree} 
\;+\; {\cal O}(\e).
\eea

Here, $F_j$ stand for such evanescent operators that are not necessary
as counterterms for the one-loop Green functions with insertions of
the physical operators $Q_i$. The matrices $\hat{c}$ and
$\hat{a}^{12}$ depend on the structure of $F_j$, but
$\hat{\gamma}^{(1)}$ does not.

\begin{figure}[htb]
\begin{center}
\includegraphics[width=5cm,angle=0]{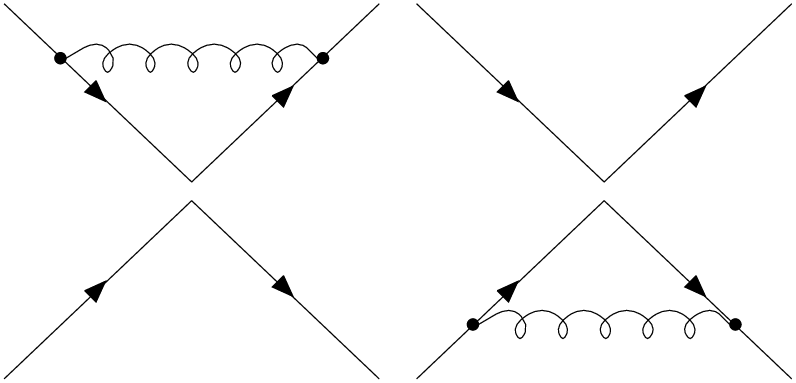}
\includegraphics[width=5cm,angle=0]{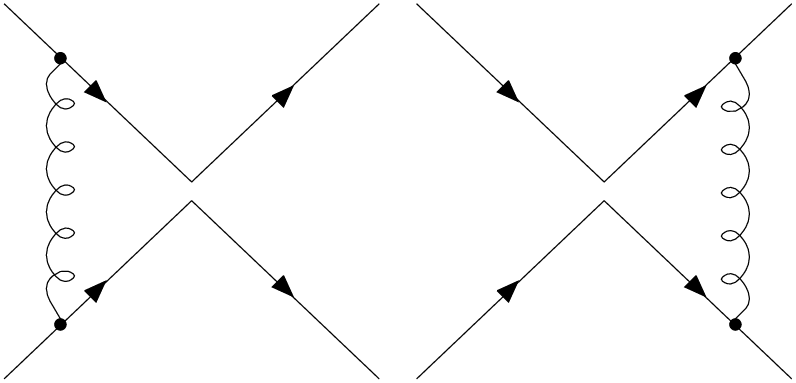}
\includegraphics[width=5cm,angle=0]{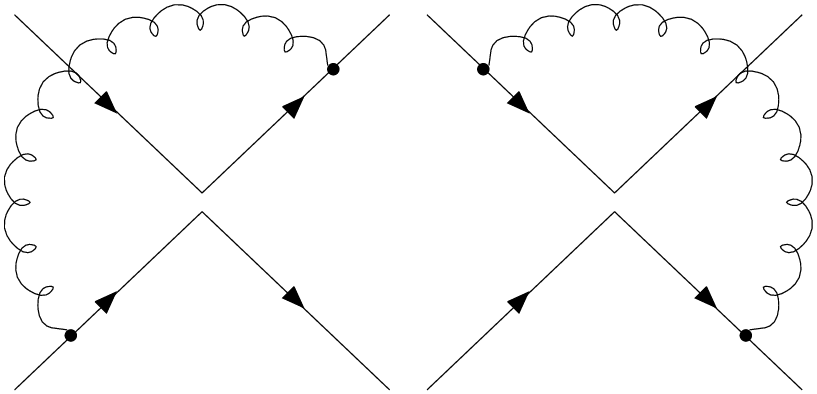}
\caption{One-loop diagrams}
\label{oneloop}
\end{center}
\end{figure}

The matrices $\hat{\gamma}^{(0)} = 2 \hat{a}^{11}$,~ $\hat{b}$ and
$\hat{c}$ in each sector are found from the one-loop $d \bar{s} \to s
\bar{d}$ diagrams presented in fig.~\ref{oneloop} with insertions of
the physical operators $Q_i$, as well as the evanescent operators
$E_k$. We calculate only the ``annihilation-type'' diagrams, i.e. we
drop all the diagrams where fermion lines connect the incoming and
outgoing particles.  Dropping such diagrams consistently at the tree
level, at one loop and (later) at two loops does not alter the final
results for the renormalization constants.

All the one- and two-loop diagrams considered in the present article
are calculated using two different methods. In both of them, a
covariant gauge-fixing term
\be \label{gauge.fix} 
{\cal L}_{gf} = -\f{1}{2 \lambda} 
(\partial^{\mu} G^a_{\mu} ) (\partial^{\nu} G^a_{\nu} )
\ee
is used, and the physical masses are set to zero. In the first method,
the external quarks are assumed to have momentum $\pm p$. In the
second method, the external momenta are set to zero, but a common mass
parameter is introduced in all the propagator denominators as IR
regulator \cite{CMM98.be}. The two methods give the same results for the
$\overline{\rm MS}$ renormalization constants. The ADMs calculated
from these renormalization constants with the help of eqs.~(\ref{gamma0})
and (\ref{gamma1}) are independent of the gauge-fixing parameter
$\lambda$.

We begin with presenting the ADM in the SLL sector, because in this
very sector our results are going to differ (at two loops) from those
of ref.~\cite{CFLMSS97}. The matrices $\hat{\gamma}^{(0)\rm SLL}$ and
$\hat{b}^{\rm SLL}$ are found to be the following:
\bea
\hat{\gamma}^{(0)\rm SLL} &=& \left( \begin{array}{ccc}
-6 N +6 +\f{6}{N} &~& \f{1}{2} -\f{1}{N} \\[1mm]
-24 -\f{48}{N}    && 2 N + 6 -\f{2}{N} 
\end{array} \right), \label{g0df2} \\[2mm]
\hat{b}^{\rm SLL} &=& \left( \begin{array}{cccc}
 0 & \f{1}{2} &           0 &        0 \\
-8 &       -8 & -\f{1}{2 N} & \f{1}{2}
\end{array} \right), 
\eea
where $N$ stands for the number of colours.

In order to find the matrix $\hat{a}^{12}$, we need to calculate
two-loop diagrams obtained from the ones in fig.~\ref{oneloop} by
including one-loop corrections on the gluon lines or adding another
gluon that couples to the open quark lines.  Of course, one-loop
diagrams with counterterm insertions need to be included, too.  All
the two-loop diagrams and the corresponding colour factors are the
same as in fig.~2 and table~2 of ref.~\cite{BW90}. However, in the
present article, we also consider additional Dirac structures
(\ref{newstr}) in the four-quark vertices.

Inserting the calculated matrix $\hat{a}^{12}$ into
eq.~(\ref{gamma1}), we obtain the two-loop ADM.  Its entries are found
to be the following:
\be \label{g1df2} \begin{array}{rcl} 
 \gamma^{(1)\rm SLL}_{11} &=& -\f{203}{6} N^2 +\f{107}{3} N +\f{136}{3}
-\f{12}{N} -\f{107}{2 N^2} +\f{10}{3} N f -\f{2}{3} f -\f{10}{3 N} f,\\[2mm]
\gamma^{(1)\rm SLL}_{12} &=& -\f{1}{36} N -\f{31}{9} +\f{9}{N} -\f{4}{N^2}
-\f{1}{18} f +\f{1}{9 N} f,\\[2mm]
\gamma^{(1)\rm SLL}_{21} &=& -\f{364}{3} N -\f{704}{3} -\f{208}{N}
-\f{320}{N^2} +\f{136}{3} f + \f{176}{3 N} f,\\[2mm]
\gamma^{(1)\rm SLL}_{22} &=& \f{343}{18} N^2 +21 N -\f{188}{9} +\f{44}{N}
+\f{21}{2 N^2} -\f{26}{9} N f -6 f +\f{2}{9 N} f, 
\end{array}
\ee
where $f$ stands for the number of active flavours. The above equation
is one of the main results of the present paper.

Proceeding analogously in the VLL sector, we reproduce the well-known
results for the one- and two-loop anomalous dimensions of the operator
$Q_1^{\rm VLL}$ \cite{BJW90}:
\be \label{gVLL.df2}
\begin{array}{rcl}
\gamma^{(0)\rm VLL} &=& 6 - \f{6}{N}, \\[2mm]
\gamma^{(1)\rm VLL} &=& -\f{19}{6} N -\f{22}{3} +\f{39}{N} -\f{57}{2 N^2} 
                 +\f{2}{3} f -\f{2}{3 N} f.
\end{array}
\ee
The matrix $\hat{b}$ in the VLL sector reads
\be
\hat{b}^{\rm VLL} = \left( \begin{array}{ccc}
-5 & -\f{1}{2 N} & \f{1}{2} \end{array} \right).
\ee

Finally, our results for the LR sector read
\bea \label{g0LR.df2}
\hat{\gamma}^{(0)\rm LR} &=& \left( \begin{array}{ccc} 
\f{6}{N} &~& 12 \\[1mm]
0 && - 6 N + \f{6}{N} \end{array} \right),
\\[2mm] \label{g1LR.df2}
\hat{\gamma}^{(1)\rm LR} &=& \left( \begin{array}{ccc} 
\f{137}{6} + \f{15}{2 N^2} - \f{22}{3N} f 
&~~&
\f{200}{3} N -\f{6}{N} -\f{44}{3} f 
\\[1mm] 
\f{71}{4} N +\f{9}{N} -2f 
&&
-\f{203}{6} N^2 +\f{479}{6} +\f{15}{2 N^2} +\f{10}{3} N f -\f{22}{3N} f 
\end{array} \right),
\\[2mm]
\hat{b}^{\rm LR} &=& \left( \begin{array}{cccccc} 
 0 & -5& -\f{1}{2 N} & \f{1}{2} & 0 & 0 \\
 0 &  0&           0 &        0 & -\f{1}{2 N} & \f{1}{2} 
\end{array} \right).
\eea
As mentioned in the introduction, all the comparisons with existing
literature are relegated to section \ref{sec:comp}.

\newsection{Current--current contributions to the ADM\\ 
of $\Delta F=1$ operators}
\label{sec:df1}

In the present section, we evaluate contributions from the
current--current diagrams to the ADM of $\Delta F=1$ operators.  For
this purpose, we choose the operators in such a manner that all the
four flavours they contain are different: $\bar{s}$, $d$, $\bar{u}$,
$c$. In such a case, the only possible diagrams are the
current--current ones.

Twenty linearly independent operators can be built out of four
different quark fields.  They can be split into 8 separate sectors,
between which there is no mixing. The operators belonging to the first
four sectors (VLL, VLR, SLR and SLL) read
\bea 
Q_1^{\rm VLL} &=& (\bar{s}^{\alpha} \gamma_{\mu}    P_L d^{ \beta})
              (\bar{u}^{ \beta} \gamma^{\mu}    P_L c^{\alpha}) ~=~ 
\tilde{Q}_{V_L V_L},
\nnb\\
Q_2^{\rm VLL} &=& (\bar{s}^{\alpha} \gamma_{\mu}    P_L d^{\alpha})
              (\bar{u}^{ \beta} \gamma^{\mu}    P_L c^{ \beta}) ~=~ 
Q_{V_L V_L},
\nnb\\[4mm]
Q_1^{\rm VLR} &=& (\bar{s}^{\alpha} \gamma_{\mu}    P_L d^{ \beta})
              (\bar{u}^{ \beta} \gamma^{\mu}    P_R c^{\alpha}) ~=~ 
\tilde{Q}_{V_L V_R},
\nnb\\
Q_2^{\rm VLR} &=& (\bar{s}^{\alpha} \gamma_{\mu}    P_L d^{\alpha})
              (\bar{u}^{ \beta} \gamma^{\mu}    P_R c^{ \beta}) ~=~ 
Q_{V_L V_R},
\nnb\\[4mm]
Q_1^{\rm SLR} &=& (\bar{s}^{\alpha}                 P_L d^{ \beta})
              (\bar{u}^{ \beta}                 P_R c^{\alpha}) ~=~ 
\tilde{Q}_{LR},
\nnb\\
Q_2^{\rm SLR} &=& (\bar{s}^{\alpha}                 P_L d^{\alpha})
              (\bar{u}^{ \beta}                 P_R c^{ \beta}) ~=~ 
Q_{LR},
\nnb\\[4mm]
Q_1^{\rm SLL} &=& (\bar{s}^{\alpha}                 P_L d^{ \beta})
              (\bar{u}^{ \beta}                 P_L c^{\alpha}) ~=~ 
\tilde{Q}_{LL},
\nnb\\
Q_2^{\rm SLL} &=& (\bar{s}^{\alpha}                 P_L d^{\alpha})
              (\bar{u}^{ \beta}                 P_L c^{\beta}) ~=~ 
Q_{LL},
\nnb\\
Q_3^{\rm SLL} &=& (\bar{s}^{\alpha} \sigma_{\mu\nu} P_L d^{ \beta})
              (\bar{u}^{ \beta} \sigma^{\mu\nu} P_L c^{\alpha}) ~=~ 
\tilde{Q}_{T_L T_L},
\nnb\\
Q_4^{\rm SLL} &=& (\bar{s}^{\alpha} \sigma_{\mu\nu} P_L d^{\alpha})
              (\bar{u}^{ \beta} \sigma^{\mu\nu} P_L c^{ \beta}) ~=~ 
Q_{T_L T_L},
\label{normal.df1}
\eea
where on the r.h.s. we have shown the notation of ref.~~\cite{CFLMSS97}.

The operators belonging to the four remaining sectors (VRR, VRL, SRL
and SRR) are obtained from the above by interchanging $P_L$ and $P_R$.
Obviously, it is sufficient to calculate the ADMs only for the VLL,
VLR, SLR and SLL sectors.  The ``mirror'' operators in the VRR, VRL,
SRL and SRR sectors will have exactly the same properties under QCD
renormalization.

The evanescent operators for the VLL, VLR, SLR and SLL sectors are
listed in appendix B. Calculation of the renormalization constants and
the ADMs proceeds along the same lines as in the previous section.
The relevant divergences in one- and two-loop diagrams in the cases
of VLL, VLR and SLR sectors are given in refs. \cite{BW90} and
\cite{BJLW93}.  For completeness we give in appendix E the
corresponding results for the SLL sector. These have not been
published so far in the NDR--$\overline{\rm MS}$ scheme.

Our final results for the $\Delta F=1$~ ADMs are as follows:
\bea
\hat{\gamma}^{(0)\rm VLL} &=& \left( \begin{array}{ccc}
-\f{6}{N} && 6 \\[1mm] 
6 && -\f{6}{N}
\end{array} \right), \label{ga0VLL} \\[2mm]
\hat{\gamma}^{(1)\rm VLL} &=& \left( \begin{array}{ccc}
-\f{22}{3} - \f{57}{2N^2} - \f{2}{3N} f &~~&
- \f{19}{6} N + \f{39}{N} +\f{2}{3}f \\[1mm] 
- \f{19}{6} N + \f{39}{N} +\f{2}{3}f &&
-\f{22}{3} - \f{57}{2N^2} - \f{2}{3N} f
\end{array} \right), \label{ga1VLL}\\[2mm]
\hat{\gamma}^{(0)\rm VLR} &=& \left( \begin{array}{ccc}
- 6N +\f{6}{N} && 0 \\[1mm] 
-6 && \f{6}{N}
\end{array} \right), \label{ga0VLR} \\[2mm]
\hat{\gamma}^{(1)\rm VLR} &=& \left( \begin{array}{ccc}
-\f{203}{6} N^2 +\f{479}{6} + \f{15}{2N^2} + \f{10}{3} N f -\f{22}{3N} f &~~~&
-\f{71}{2} N - \f{18}{N} +4f \\[1mm] 
- \f{100}{3} N + \f{3}{N} + \f{22}{3} f &&
\f{137}{6} + \f{15}{2N^2} - \f{22}{3N} f
\end{array} \right), \label{ga1VLR} \\[2mm]
\hat{\gamma}^{(0)\rm SLR} &=& \left( \begin{array}{ccc}
\f{6}{N} && -6 \\[1mm] 
0 && - 6N+ \f{6}{N} 
\end{array} \right), \label{ga0SLR} \\[2mm]
\hat{\gamma}^{(1)\rm SLR} &=& \left( \begin{array}{ccc}
\f{137}{6} + \f{15}{2N^2} - \f{22}{3N} f &~~~&
-\f{100}{3} N + \f{3}{N} + \f{22}{3} f \\[1mm] 
- \f{71}{2} N - \f{18}{N} + 4f &&
-\f{203}{6} N^2 +\f{479}{6} + \f{15}{2N^2} + \f{10}{3} N f -\f{22}{3N} f
\end{array} \right), \label{ga1SLR} \\[2mm]
\hat{\gamma}^{(0)\rm SLL} &=& \left( \begin{array}{cccc}
\f{6}{N} & -6 & \f{N}{2}-\f{1}{N} & \f{1}{2} \\ 
0 & - 6N+\f{6}{N} & 1 & -\f{1}{N} \\ 
-\f{48}{N} + 24N & 24 & -\f{2}{N} - 4N & 6 \\ 
48  & -\f{48}{N} & 0 & 2N-\f{2}{N} 
\end{array} \right), \label{ga0SLL}
\eea
\bea 
\gamma^{(1)\rm SLL}_{11} &=& \begin{array}{l}
-\f{N^2}{2} + \f{148}{3} - \f{107}{2N^2} - 2 N f - \f{10}{3N} f, 
\end{array}\nnb\\
\gamma^{(1)\rm SLL}_{12} &=& \begin{array}{l}
-\f{178}{3} N + \f{64}{N} + \f{16}{3} f, 
\end{array}\nnb\\
\gamma^{(1)\rm SLL}_{13} &=& \begin{array}{l}
\f{107}{36} N^2 -\f{71}{18} - \f{4}{N^2} - \f{1}{18} N f + \f{f}{9N}, 
\end{array}\nnb\\
\gamma^{(1)\rm SLL}_{14} &=& \begin{array}{l}
-\f{109}{36} N + \f{8}{N} -\f{f}{18},
\end{array}\nnb\\
\gamma^{(1)\rm SLL}_{21} &=& \begin{array}{l}
-26N + \f{104}{N}, 
\end{array}\nnb\\
\gamma^{(1)\rm SLL}_{22} &=& \begin{array}{l}
-\f{203}{6} N^2 + \f{28}{3} - \f{107}{2N^2} + \f{10}{3} N f - \f{10}{3N} f, 
\end{array}\nnb\\
\gamma^{(1)\rm SLL}_{23} &=& \begin{array}{l}
\f{89}{18} N + \f{2}{N} -\f{1}{9} f, 
\end{array}\nnb\\
\gamma^{(1)\rm SLL}_{24} &=& \begin{array}{l}
-\f{53}{18} - \f{4}{N^2} + \f{1}{9N} f, 
\end{array}\nnb\\
\gamma^{(1)\rm SLL}_{31} &=& \begin{array}{l}
\f{676}{3} N^2 -\f{1880}{3} - \f{320}{N^2} - \f{88}{3} N f + \f{176}{3N} f, 
\end{array}\nnb\\
\gamma^{(1)\rm SLL}_{32} &=& \begin{array}{l}
\f{820}{3} N + \f{448}{N} -\f{88}{3} f, 
\end{array}\nnb\\
\gamma^{(1)\rm SLL}_{33} &=& \begin{array}{l}
-\f{257}{18} N^2 -\f{116}{9} + \f{21}{2N^2} + \f{22}{9} N f + \f{2}{9N} f, 
\end{array}\nnb\\
\gamma^{(1)\rm SLL}_{34} &=& \begin{array}{l}
\f{50}{3} N -\f{8}{3} f, 
\end{array}\nnb\\
\gamma^{(1)\rm SLL}_{41} &=& \begin{array}{l}
\f{488}{3} N + \f{416}{N} -\f{176}{3} f, 
\end{array}\nnb\\
\gamma^{(1)\rm SLL}_{42} &=& \begin{array}{l}
-\f{776}{3} - \f{320}{N^2} + \f{176}{3N} f, 
\end{array}\nnb\\
\gamma^{(1)\rm SLL}_{43} &=& \begin{array}{l}
\f{22}{3} N  - \f{40}{N}  +\f{8}{3} f, 
\end{array}\nnb\\
\gamma^{(1)\rm SLL}_{44} &=& \begin{array}{l}
\f{343}{18} N^2 + \f{28}{9} + \f{21}{2N^2} - \f{26}{9} N f + \f{2}{9N} f. 
\end{array}\label{ga1SLL} 
\eea

Equation (\ref{ga1SLL}) is one of the main results of this work.

The careful reader has already noticed that the following equalities
hold up to~ {\scriptsize ${\cal O}$}$(\al^2)$:
\be \label{fierz.rel}
\begin{array}{ccccc}
\gamma_{11}^{\rm VLL} = \gamma_{22}^{\rm VLL},
& \hspace{2cm} &
\gamma_{12}^{\rm VLL} = \gamma_{21}^{\rm VLL},
& \hspace{2cm} &
\gamma_{11}^{\rm VLR} = \gamma_{22}^{\rm SLR}, \\[2mm]
\gamma_{22}^{\rm VLR} = \gamma_{11}^{\rm SLR}, 
&&
\gamma_{12}^{\rm VLR} = \gamma_{21}^{\rm SLR},
&&
\gamma_{21}^{\rm VLR} = \gamma_{12}^{\rm SLR}.
\end{array}
\ee
At one loop, these equalities are a consequence of the Fierz
identities 
\bea
(\gamma_{\mu} P_L)_{ij} (\gamma^{\mu} P_L)_{kl} &=& -
(\gamma_{\mu} P_L)_{il} (\gamma^{\mu} P_L)_{kj}, \label{fVLL}\\
(\gamma_{\mu} P_L)_{ij} (\gamma^{\mu} P_R)_{kl} 
&=& 2 (P_R)_{il} (P_L)_{kj}, \label{fLR}
\eea
as well as the flavour- and chirality-blind character of QCD
interactions.  Since the Fierz identities are satisfied in four
spacetime dimensions only, the relations (\ref{fierz.rel}) could be
potentially broken at two loops in the NDR--$\overline{\rm MS}$ scheme.
Surprisingly, they are not.\footnote{
  In section 4, where the penguin diagrams are considered, no
  invariance under Fierz rearrangement is observed at two loops for
  the operators with VLL Dirac structure.  A detailed
  discussion of this fact can be found in ref.~\cite{BJLW93}.}

On the contrary, analogous relations {\em are} broken at two loops in
the SLL sector.  Because of the Fierz relations (\ref{fSLL}), the one-loop
matrix $\hat{\gamma}^{(0)\rm SLL}$ must satisfy the following identity
(cf.  eqs.~(9) and (10) of ref.~\cite{CFLMSS97}):
\be
\hat{\gamma}^{(0)\rm SLL} = \hat{\cal F} \hat{\gamma}^{(0)\rm SLL} \hat{\cal F}
\ee
with 
\be \hat{\cal F} = \left( \begin{array}{cccc} 
0 & -\f{1}{2} & 0 & \f{1}{8} \\
-\f{1}{2} & 0 & \f{1}{8} & 0 \\
0 & 6 & 0 & \f{1}{2} \\
6 & 0 & \f{1}{2} & 0 
\end{array} \right).
\ee
No similar relation holds for $\hat{\gamma}^{(1)\rm SLL}$ in the
NDR--$\overline{\rm MS}$ scheme. As it has already been said, this
is not surprising, because the Fierz identities are not true in $D\neq
4$ dimensions.

It is unclear to us whether the symmetries (\ref{fierz.rel}) for the
VLL, VLR and SLR sectors are preserved at two loops in the
NDR--$\overline{\rm MS}$ scheme only by coincidence, or if there is
some reason beyond this. As we shall see in section~\ref{sec:transf},
this question is related to the properties of one-loop matrix elements
of certain evanescent operators.

\newsection{Penguin contributions to the ADM\\ of $\Delta F=1$ operators}
\label{sec:penguin}

In the present section, we shall describe additional contributions to
the ADM of $\Delta F=1$ operators that are due to penguin diagrams.
Such contributions may arise only when the operators contain one
quark-antiquark pair of the same flavour.

For definiteness, let us consider $\Delta S=1$ operators. In the SM
analysis of ref.~\cite{BJLW93}, 10 such operators were considered\footnote{
  Our operators here differ from the ones in ref.~\cite{BJLW93} by a
  global normalization factor of 4. Of course, it does not affect their
  ADM.  The factor of 4 can be absorbed into the global normalization
  factor of the effective Lagrangian, as the first ratio on the
  r.h.s. of eq.~(\ref{Leff}). In this case, the Wilson coefficients 
  of our operators are exactly the same as those in ref.~\cite{BJLW93}.}
\bea
Q_1 &=& (\bar{s}^{\alpha} \gamma_{\mu} P_L u^{\beta })
        (\bar{u}^{\beta } \gamma^{\mu} P_L d^{\alpha}),\nnb\\
Q_2 &=& (\bar{s}^{\alpha} \gamma_{\mu} P_L u^{\alpha})
        (\bar{u}^{\beta } \gamma^{\mu} P_L d^{\beta }),\nnb\\
Q_3 &=& (\bar{s}^{\alpha} \gamma_{\mu} P_L d^{\alpha}) \sum_q
        (\bar{q}^{\beta } \gamma^{\mu} P_L q^{\beta }),\nnb\\
Q_4 &=& (\bar{s}^{\alpha} \gamma_{\mu} P_L d^{\beta }) \sum_q
        (\bar{q}^{\beta } \gamma^{\mu} P_L q^{\alpha}),\nnb\\
Q_5 &=& (\bar{s}^{\alpha} \gamma_{\mu} P_L d^{\alpha}) \sum_q
        (\bar{q}^{\beta } \gamma^{\mu} P_R q^{\beta }),\nnb\\
Q_6 &=& (\bar{s}^{\alpha} \gamma_{\mu} P_L d^{\beta }) \sum_q
        (\bar{q}^{\beta } \gamma^{\mu} P_R q^{\alpha}),\nnb\\
Q_7 &=& \f{3}{2} 
        (\bar{s}^{\alpha} \gamma_{\mu} P_L d^{\alpha}) \sum_q e_q
        (\bar{q}^{\beta } \gamma^{\mu} P_R q^{\beta }),\nnb\\
Q_8 &=& \f{3}{2}
        (\bar{s}^{\alpha} \gamma_{\mu} P_L d^{\beta }) \sum_q e_q
        (\bar{q}^{\beta } \gamma^{\mu} P_R q^{\alpha}),\nnb\\
Q_9 &=& \f{3}{2}
        (\bar{s}^{\alpha} \gamma_{\mu} P_L d^{\alpha}) \sum_q e_q
        (\bar{q}^{\beta } \gamma^{\mu} P_L q^{\beta }),\nnb\\
Q_{10} &=& \f{3}{2}
        (\bar{s}^{\alpha} \gamma_{\mu} P_L d^{\beta }) \sum_q e_q
        (\bar{q}^{\beta } \gamma^{\mu} P_L q^{\alpha}).
\label{SM.ops}
\eea
Their one- and two-loop ADMs, including current--current and penguin
diagrams, can be found in appendices A and B of ref.~\cite{BJLW93}.
They were also obtained in ref.~\cite{CFMR94}. The same results hold
for the mirror copies of the SM operators, i.e.  for the operators
obtained from the ones in eq.~(\ref{SM.ops}) by $P_L \leftrightarrow
P_R$ interchange.

Beyond SM, new linearly independent operators appear. Their Dirac
structures are as in eq.~(\ref{normal.df1}). Our aim is to find a
minimal set of linearly independent new operators. In the process of
identifying these operators, we shall use four-dimensional Dirac
algebra, including the Fierz relations (\ref{fSLL}), (\ref{fVLL}) and
(\ref{fLR}). It turns out that only 3 additional operators (and their
mirror copies) undergo mixing via penguin diagrams into other
four-quark operators in eq.~(\ref{SM.ops}). These are
\bea
Q_{11} &=& (\bar{s}^{\alpha} \gamma_{\mu} P_L d^{\alpha})
    \left[ (\bar{d}^{\beta } \gamma^{\mu} P_L d^{\beta })
          +(\bar{s}^{\beta } \gamma^{\mu} P_L s^{\beta }) \right],\nnb\\
Q_{12} &=& (\bar{s}^{\alpha} \gamma_{\mu} P_L d^{\beta })
    \left[ (\bar{d}^{\beta } \gamma^{\mu} P_R d^{\alpha})
          +(\bar{s}^{\beta } \gamma^{\mu} P_R s^{\alpha}) \right],\nnb\\
Q_{13} &=& (\bar{s}^{\alpha} \gamma_{\mu} P_L d^{\alpha})
    \left[ (\bar{d}^{\beta } \gamma^{\mu} P_R d^{\beta })
          +(\bar{s}^{\beta } \gamma^{\mu} P_R s^{\beta }) \right].
\label{additional}
\eea
The remaining elements of the operator basis can be chosen in such a
manner that massless penguin diagrams with their insertions vanish.
The first three of the remaining operators have the structure of
$Q_{11}$, ..., $Q_{13}$, but with a relative minus sign between the
two terms. The next two have the structure of $Q_5$ and $Q_6$,
but the sum over flavour-conserving currents is replaced by a
difference between the analogous $u$-quark and $c$-quark currents.
Their mirror copies have to be included, as well.  Further operators
have the SLL and SRR Dirac structures as in eq.~(\ref{newstr}), or
they have the form
\bea
(\bar{s}^{\alpha} P_{L,R} d^{\beta })
(\bar{q}^{\beta } P_{R,L} q^{\alpha}),
\label{no.penguin.1}\\
(\bar{s}^{\alpha} P_{L,R} d^{\alpha})
(\bar{q}^{\beta } P_{R,L} q^{\beta })
\label{no.penguin.2}
\eea
where $q$ has flavour {\it different} from $s$ and $d$.  It is
straightforward to convince oneself that we have not missed any
linearly independent $\Delta S=1$ operator in the above
considerations.

Massless penguin diagrams with insertions of the operators
(\ref{newstr}), (\ref{no.penguin.1}) and (\ref{no.penguin.2}) vanish,
because
\be
{\rm Tr}(S_{\rm odd} P_{L,R}) = 0 \hspace{2cm} {\rm and} \hspace{2cm} 
P_{L,R} S_{\rm odd} P_{L,R} = 0,
\ee
where $S_{\rm odd}$ is a product of an odd number of Dirac
$\gamma$-matrices.  For dimensional reasons, only massless penguin
diagrams can cause mixing into other four-quark operators. This means
that all the $\Delta S=1$ operators, except for $Q_1$, ..., $Q_{13}$
and their mirror copies, mix only due to current--current diagrams, i.e.
their ADMs are identical to the ones we have already calculated in
sections~\ref{sec:df2} and \ref{sec:df1}.

At the two-loop level, a complication arises because generally the
Fierz relations could be broken in $D\not=4$ dimensions.
Consequently, our use of these relations in the identification of
linearly independent operators could be put in question.  However, as
we have already discussed in section 2 and will elaborate in section
5, this complication can be avoided by introducing appropriate
evanescent operators that vanish in four dimensions by Fierz
identities.  This allows us to restrict the basis of new physical
operators (undergoing penguin mixing) to the one in
eq.~(\ref{additional}), even at the two-loop level.

The introduction of evanescent operators that vanish in four
dimensions by Fierz identities turns out to have no effect on the
two-loop ADM in the case of the operators with VLR and SLR structures,
because the Fierz identity (\ref{fLR}) remains valid at two loops in
the NDR--$\overline{\rm MS}$ scheme, even if the penguin insertions
are considered \cite{BJLW93}. On the other hand, as pointed out in
ref.~\cite{BJLW93}, the Fierz identity (\ref{fVLL}) is broken at
two loops in the NDR--$\overline{\rm MS}$ scheme through penguin
diagrams, although it remains valid for current--current diagrams.  As
a result, the mixing of the operator
\be
Q_{11}^{\prime} = (\bar{s}^{\alpha} \gamma_{\mu} P_L d^{\beta})
    \left[ (\bar{d}^{\beta } \gamma^{\mu} P_L d^{\alpha })
          +(\bar{s}^{\beta } \gamma^{\mu} P_L s^{\alpha }) \right]
\label{addprime}
\ee
with the operators in eq.~(\ref{SM.ops}), through penguin diagrams,
differs from the one of $Q_{11}$ at the two-loop level. This can be
easily verified by using the results of ref.~\cite{BJLW93}. As
$Q_{11}^{\prime}=Q_{11}$ in $D=4$ dimensions due to the Fierz identity
(\ref{fVLL}), $Q_{11}^{\prime}$ was not included in the basis
(\ref{additional}). By working with $Q_{11}$ and the evanescent
operator $Q_{11}^{\prime}-Q_{11}$, the explicit appearance of
$Q_{11}^{\prime}$ can be avoided at any number of loops, so that the
basis (\ref{additional}) remains unchanged.

The above discussion implies that the only additional ADMs we need
to find in the present section are:
\begin{itemize}
\item{} The 3$\times$3 matrix $\hat{\gamma}_{cc}$ describing the
  mixing of $Q_{11}$, ..., $Q_{13}$ among themselves.
\item{} The 3$\times$4 matrix $\hat{\gamma}_p$ describing the mixing
  of $Q_{11}$, ..., $Q_{13}$ into $Q_3$, ..., $Q_6$ via penguin
  diagrams. (Only $Q_3$, ..., $Q_6$ are generated by massless QCD
  penguin diagrams with four-quark operator insertions.)
\end{itemize}
The matrix $\hat{\gamma}_{cc}$ is given by current--current diagrams
only. It takes the form
\be \label{gammacc}
\hat{\gamma}_{cc} = \left( \begin{array}{cc}
\gamma^{\rm VLL}_{\Delta F = 2} & 0 \\[2mm]
0 & \hat{\gamma}^{\rm VLR}_{\Delta F = 1}
\end{array} \right)
\ee
with $\gamma^{\rm VLL}_{\Delta F = 2}$ and $\hat{\gamma}^{\rm VLR}_{\Delta F =
  1}$ taken from eqs.~(\ref{gVLL.df2}), (\ref{ga0VLL}) and
(\ref{ga1VLL}).

The matrix ~$\hat{\gamma}_p = \hat{\gamma}^{(0)}_p + {\displaystyle
  \f{\al}{4\pi}} \hat{\gamma}^{(1)}_p + ...$~ that originates from
penguin diagrams can be easily extracted from sections 3.2 and 5.3
of ref.~\cite{BJLW93}.  We find\footnote{
The four integers marked in red in eq.~(\ref{ga1peng}) have been
corrected with respect to the first ({\tt v1}) arXiv version of the
current paper. They are now in agreement with
ref.~\cite{Morell:2024aml} where a mistake in our original
determination of $\hat{\gamma}^{(1)}_p$ was pointed out. In the
current version of the article, we include an extra appendix~F where
our extraction of $\hat{\gamma}^{(0)}_p$ and $\hat{\gamma}^{(1)}_p$
from the results of ref.~\cite{BJLW93} is outlined.}
\bea \label{ga0peng}
\hat{\gamma}^{(0)}_p &=& 
\begin{array}{c} \left(\f{4}{3}, \f{4}{3}, 0 \right)^T \times
\left(-\f{1}{N},1,-\f{1}{N},1 \right), \end{array}\\[2mm]
\hat{\gamma}^{(1)T}_p &=& \left( \begin{array}{ccccc}
6N -\f{64}{27} -\f{{\g 4}}{3N} + \f{172}{27N^2} && 
-\f{112}{27} - \f{356}{27N^2} &&
-6N +\f{40}{3N}\\[2mm]
\f{352}{27}N -\f{{\g 14}}{3} -\f{460}{27N} &&
-\f{32}{27} N +\f{500}{27N} &&
-\f{22}{3}\\[2mm]
-6N -\f{244}{27} +\f{{\g 32}}{3N} -\f{188}{27N^2} &&
\f{140}{27} + \f{148}{27N^2} &&
6N +\f{4}{3N}\\[2mm]
\f{172}{27} N -\f{{\g 14}}{3} +\f{260}{27N} &&
\f{220}{27} N -\f{508}{27N} &&
-\f{22}{3}
\end{array} \right).
\label{ga1peng}
\eea

The above discussion changes very little in the case of $\Delta F=1$
operators, in which $F$ is the up-type flavour. Similarly to the
$\Delta S=1$ case, all the contributions from penguin diagrams can be
easily extracted from ref.~~\cite{BJLW93}.

\newsection{Transformation of the Wilson coefficients\\ to the RI scheme}
\label{sec:RI}

The ADMs calculated in the present work are given in the
NDR--$\overline{\rm MS}$ scheme that is most convenient for
perturbative calculations. However, after the Wilson coefficients are
evolved with the help of RGE (\ref{RGE}) down to a low energy scale,
it might be necessary to transform them to another scheme that is more
appropriate for non-perturbative calculations of hadronic matrix
elements \cite{BMRGGLM00}. One such scheme is the so-called
Regularization-Independent (RI) scheme (originally called the MOM
scheme) used in ref.~\cite{CFLMSS97}. Below, we shall give relations
between the NDR--$\overline{\rm MS}$-renormalized and
RI-renormalized Wilson coefficients of all the operators considered
in sections \ref{sec:df2} and \ref{sec:df1}.

For completeness, we begin with the definition of the RI scheme. For
the massless quark propagator, the renormalization condition can be
written as
\be \label{RI.cond.prop}
\f{i}{4} \left[ \gamma^{\rho} \f{\partial}{\partial p^{\rho}}
S(p)_{\scs R}^{-1} \right]_{p^2=-\mu^2}= 1,
\ee
where $\mu$ is the subtraction scale. A simple one-loop calculation is
necessary to verify that the renormalized inverse propagator in the RI
scheme reads
\be
S(p)_{\scs R}^{-1} = -i \slash p \left[ 1 - \f{\al}{4 \pi} C_F \lambda 
\left( \f{1}{2} - \ln \f{-p^2}{\mu^2} \right) \right] + {\cal O}(\al^2),
\ee
where $C_F = \f{N^2-1}{2N}$ and $\lambda$ is the gauge-fixing
parameter (cf. eq.~(\ref{gauge.fix})). In dimensional regularization,
the corresponding quark wave-function renormalization constant reads
\be \label{Zq.RI}
Z_q^{\rm RI}=1-\f{\al}{4\pi}C_F\lambda\left(\f{1}{\e}-\gamma+\ln(4\pi)+\f{1}{2}\right),
\ee
provided the subtraction scale $\mu$ is identified with the standard
$\overline{\rm MS}$ renormalization scale.

Conditions similar to eq.~(\ref{RI.cond.prop}) are imposed on
renormalized matrix elements of the operators (\ref{normal}) and
(\ref{normal.df1}) among four external quarks with the same momentum
$p$. The quarks are assumed to be massless here. For the $\Delta F=2$
operators, such matrix elements have the following form
\bea
\me{Q_1^{\rm VLL}}_{\scs R} &=& A_{11}^{\rm VLL}(p^2) \; \me{Q_1^{\rm VLL}}_{\rm tree} ~+~
B_{11}^{\rm VLL}(p^2) \; p^{\mu} p_{\nu}\; \me{
(\bar{s}^{\alpha} \gamma_{\mu}     P_L d^{\alpha})
(\bar{s}^{ \beta} \gamma^{\nu}     P_L d^{ \beta})}_{\rm tree},\nnb\\
\me{Q_1^{\rm LR}}_{\scs R} &=& A_{11}^{\rm LR}(p^2) \; \me{Q_1^{\rm LR}}_{\rm tree} ~+~
B_{11}^{\rm LR}(p^2) \; p^{\mu} p_{\nu}\; \me{
(\bar{s}^{\alpha} \gamma_{\mu}     P_L d^{\alpha})
(\bar{s}^{ \beta} \gamma^{\nu}     P_R d^{ \beta})}_{\rm tree} \nnb\\ 
&+& A_{12}^{\rm LR}(p^2) \; \me{Q_2^{\rm LR}}_{\rm tree} ~+~
B_{12}^{\rm LR}(p^2) \; p^{\mu} p_{\nu}\; \me{
(\bar{s}^{\alpha} \sigma_{\mu\rho} P_L d^{\alpha})
(\bar{s}^{ \beta} \sigma^{\nu\rho} P_R d^{ \beta})}_{\rm tree},\nnb\\ 
\me{Q_2^{\rm LR}}_{\scs R} &=& A_{21}^{\rm LR}(p^2) \; \me{Q_1^{\rm LR}}_{\rm tree} ~+~
B_{21}^{\rm LR}(p^2) \; p^{\mu} p_{\nu}\; \me{
(\bar{s}^{\alpha} \gamma_{\mu}     P_L d^{\alpha})
(\bar{s}^{ \beta} \gamma^{\nu}     P_R d^{ \beta})}_{\rm tree} \nnb\\ 
&+& A_{22}^{\rm LR}(p^2) \; \me{Q_2^{\rm LR}}_{\rm tree} ~+~
B_{22}^{\rm LR}(p^2) \; p^{\mu} p_{\nu}\; \me{
(\bar{s}^{\alpha} \sigma_{\mu\rho} P_L d^{\alpha})
(\bar{s}^{ \beta} \sigma^{\nu\rho} P_R d^{ \beta})}_{\rm tree},\nnb\\ 
\me{Q_1^{\rm SLL}}_{\scs R} &=& A_{11}^{\rm SLL}(p^2) \; \me{Q_1^{\rm SLL}}_{\rm tree} 
                         +  A_{12}^{\rm SLL}(p^2) \; \me{Q_2^{\rm SLL}}_{\rm tree},\nnb\\ 
\me{Q_2^{\rm SLL}}_{\scs R} &=& A_{21}^{\rm SLL}(p^2) \; \me{Q_1^{\rm SLL}}_{\rm tree}
                         +  A_{22}^{\rm SLL}(p^2) \; \me{Q_2^{\rm SLL}}_{\rm tree}.
\label{melem}
\eea
The formfactors $B_{ij}(p^2)$ originate from UV-finite parts of
Feynman diagrams and are scheme-independent.  Note that in all the
matrix elements multiplied by $B_{ij}(p^2)$, only colour-singlet quark
currents occur.  Colour-octet currents are removed from these terms
with the help of the following Fierz identities (which are independent from
the ones in eqs.~(\ref{fSLL}), (\ref{fVLL}) and (\ref{fLR})):
\bea
(\slash p P_L)_{ij}(\slash p P_L)_{kl} &=& 
(\slash p P_L)_{il}(\slash p P_L)_{kj}
-\f{1}{2} p^2 (\gamma_{\mu} P_L)_{il}(\gamma^{\mu} P_L)_{kj},\\
(\slash p P_L)_{ij}(\slash p P_R)_{kl} &=& 
\f{1}{2} p^{\mu} p_{\nu} (\sigma_{\mu\rho} P_R)_{il}(\sigma^{\nu\rho} P_L)_{kj}
+ \f{1}{2} p^2 (P_R)_{il} (P_L)_{kj},\\ 
p^{\mu} p_{\nu} (\sigma_{\mu\rho} P_L)_{ij}(\sigma^{\nu\rho} P_R)_{kl} &=&
2 (\slash p P_R)_{il}(\slash p P_L)_{kj} 
- \f{1}{2} p^2 (\gamma_{\mu} P_R)_{il} (\gamma^{\mu} P_L)_{kj}.
\eea
No $B_{ij}$ formfactors occur in the SLL sector thanks to the
four-dimensional identity
\be
p^{\mu} p_{\nu} (\sigma_{\mu\rho} P_L)_{ij}(\sigma^{\nu\rho} P_L)_{kl} =
\f{1}{4} p^2 (\sigma_{\mu\nu} P_L)_{ij} (\sigma^{\mu\nu} P_L)_{kl}.
\ee

The RI renormalization condition reads
\be \label{RI.cond.oper}
A_{ij}(-\mu^2) \;-\; \omega \mu^2 B_{ij}(-\mu^2) = \delta_{ij},
\ee
with 
\be
\omega = \left\{ \begin{array}{cl}
\f{1}{4} & {\rm for}~~ B_{11}^{\rm VLL},~ B_{11}^{\rm LR} {\rm ~~and~~} B_{21}^{\rm LR},\\[2mm]
0 & {\rm otherwise.}
\end{array} \right.
\ee

The renormalization condition (\ref{RI.cond.oper}) can be equivalently
written as
\be \label{eff.RI.cond.oper}
A_{ij}^{\rm effective}(p^2 = -\mu^2) = \delta_{ij},
\ee
with $A_{ij}^{\rm effective}$ obtained from eqs.~(\ref{melem}) by
making the following {\it ad hoc} replacements
\bea
(\slash p P_L) \otimes (\slash p P_{L,R})  &\to& 
\f{1}{4} p^2 (\gamma_{\mu} P_L) \otimes (\gamma^{\mu} P_{L,R}), \nnb\\
p^{\mu} p_{\nu} (\sigma_{\mu\rho} P_{L,R}) \otimes (\sigma^{\nu\rho} P_{R,L}) &\to& 0.
\label{replac} 
\eea

In the case of $\Delta F=1$ operators, the general structure of
one-loop matrix elements is similar to that in eq.~(\ref{melem}), but the
number of formfactors is larger, because operators with colour-octet
currents are now linearly independent. The matrix elements can be
written as
\be
\me{Q_i}_{\scs R} = A_{ij}^{\rm effective}(p^2) \me{Q_j}_{\rm tree} + N_i,
\ee
where $N_i$ vanish under the replacements (\ref{replac}).  The RI
renormalization condition then has the same form as in
eq.~(\ref{eff.RI.cond.oper}).

In each of the sectors, the RI-renormalized Wilson coefficients can
be obtained from the NDR-$\overline{\rm MS}$--renormalized ones with
the help of the following relation
\be \label{coeff.transf}
\vec{C}^{\rm RI}(\mu) = \left( 1 -\f{\al(\mu)}{4\pi} 
\Delta \hat{r}^T_{\overline{\rm MS}\to RI}(\mu) \right) \vec{C}^{\overline{\rm MS}}(\mu)
+ {\cal O}\left(\al^2\right),
\ee
where
\be \label{dr0}
\left[ \Delta r_{\overline{\rm MS}\to RI}(\mu) \right]_{ij}
= \f{4 \pi}{\al(\mu)} \left[ A_{ij}^{\rm RI}(p^2)- A_{ij}^{\overline{\rm MS}}(p^2) \right].
\ee
The above relations can be easily derived from the fact that the
renormalized matrix element of the whole effective Hamiltonian is
scheme-independent, i.e.
\be \label{RI.eq.MS}
\vec{C}^{\rm RI}(\mu) \me{\vec{Q}(\mu,p^2)}^{\rm RI}
= \vec{C}^{\overline{\rm MS}}(\mu) \me{\vec{Q}(\mu,p^2)}^{\overline{\rm MS}}.
\ee
Again, the RI subtraction scale and the standard ${\overline{\rm MS}}$
renormalization scale have been tacitly identified. The external
states must be the same in eq.~(\ref{RI.eq.MS}). Consequently, the
RI-scheme renormalization constant (\ref{Zq.RI}) must be used for
external quark lines in $A_{ij}^{\overline{\rm MS}}(p^2)$ that enters
into eq.~(\ref{dr0}).

The dependence on $p^2$ and the explicit dependence on $\mu$ cancels
out in $\Delta \hat{r}_{\overline{\rm MS}\to RI}$~ (\ref{dr0}).
However, one should not forget that this matrix depends on the
gauge-fixing parameter $\lambda$ that is, in turn, $\mu$-dependent.

Once the RI renormalization conditions have been specified, finding
the explicit form of $\Delta \hat{r}_{\overline{\rm MS}\to RI}$ is
only a matter of a straightforward one-loop computation. Our results
for the $\Delta F=2$ operators are as follows:
\mathindent0cm
\bea
\nnb\Delta r_{\overline{\rm MS}\rightarrow {\rm RI}}^{\rm VLL} &=& 
 \begin{array}{c}
 7-\f{7}{N}-12 \lo+\f{12 \lo}{N} + \lambda \left(\f32-\f3{2 N} - 4\lo +\f{4\lo}{N}\right),
 \end{array}\\[2mm]
\nnb\Delta \hat{r}_{\overline{\rm MS}\rightarrow {\rm RI}}^{\rm LR} &=& 
 \left(\begin{array}{cc}
      \f{2}{N}+\f{2\lo}{N}+\lambda\left(\f{1}{2 N}+\f{2\lo}{N}\right) &
      4+4\lo+\lambda\left(1+4\lo\right) \\
      -1+\lo-\lambda\left(\f12-\lo\right) &
      -4 N+\f{2}{N}+\f{2 \lo}{N} - \lambda\left(\f{3 N}{2}-\f{1}{2 N}-\f{2\lo}{N}\right)
       \end{array}\right),\\[2mm]
\nnb\left[\Delta r_{\overline{\rm MS}\rightarrow {\rm RI}}^{\rm SLL}\right]_{11} &=&
 \begin{array}{c}
 -4 N+7+\f5N-4\lo+\f{2\lo}N+\lambda\left(\f12+\f1{2 N}-\f{3 N}{2}+\f{2\lo}N\right),
 \end{array}\\
\nnb\left[\Delta r_{\overline{\rm MS}\rightarrow {\rm RI}}^{\rm SLL}\right]_{12} &=&
 \begin{array}{c}
 \f{5}{12}-\f{13}{12 N} -\f{2 \lo}{3}+\f{5\lo}{6 N} +
 \lambda\left(\f{5}{24}-\f{1}{6 N}-\f{\lo}3+\f{\lo}{6 N}\right),
 \end{array}\\
\nnb\left[\Delta r_{\overline{\rm MS}\rightarrow {\rm RI}}^{\rm SLL}\right]_{21} &=&
 \begin{array}{c}
 4-\f{12}N-32\lo+\f{40\lo}N-\lambda\left(2+\f8N+16\lo-\f{8\lo}N\right),
 \end{array}\\
\left[\Delta r_{\overline{\rm MS}\rightarrow {\rm RI}}^{\rm SLL}\right]_{22} &=&
 \begin{array}{c}
 \f73-\f5{3 N}-\f{28 \lo}3+\f{26\lo}{3 N} + 
 \lambda\left(\f{N}{2}+\f76-\f5{6 N}-\f{8 \lo}3+\f{10\lo}{3 N}\right).
 \nnb
 \end{array}
\eea
In the $\Delta F=1$ case, we find
\bea
\nnb\Delta \hat{r}_{\overline{\rm MS}\rightarrow {\rm RI}}^{\rm VLL} &=&
 \left(\begin{array}{cc}
      -\f{7}{N}+\f{12\lo}{N} -\lambda\left(\f3{2 N}-\f{4\lo}{N}\right) &
      7-12\lo+\lambda\left(\f32-4\lo\right)\\
      7-12\lo+\lambda\left(\f32-4\lo\right) &
      -\f{7}{N}+\f{12\lo}{N} -\lambda\left(\f3{2 N}-\f{4\lo}{N}\right)
       \end{array}\right),\\[2mm]
\nnb\Delta \hat{r}_{\overline{\rm MS}\rightarrow {\rm RI}}^{\rm VLR} &=&
 \left(\begin{array}{cc}
      -4 N+\f2N+\f{2\lo}{N}-\lambda\left(\f{3 N}2-\f1{2 N}-\f{2 \lo}N\right) &
      2-2\lo +\lambda\left(1-2\lo\right) \\
      -2-2\lo -\lambda\left(\f12+2\lo\right) &
      \f2N+\f{2\lo}{N}+\lambda\left(\f1{2 N}+\f{2 \lo}N\right)
      \end{array}\right),\\[2mm]
\nnb\Delta \hat{r}_{\overline{\rm MS}\rightarrow {\rm RI}}^{\rm SLR} &=&
 \left(\begin{array}{cc}
      \f2N+\f{2\lo}{N}+\lambda\left(\f1{2 N}+\f{2 \lo}N\right) &
      -2-2\lo -\lambda\left(\f12+2\lo\right) \\
      2-2\lo +\lambda\left(1-2\lo\right) &
      -4 N+\f2N+\f{2\lo}{N}-\lambda\left(\f{3 N}2-\f1{2 N}-\f{2 \lo}N\right)
      \end{array}\right),\\[2mm]
\nnb\left[\Delta r_{\overline{\rm MS}\rightarrow {\rm RI}}^{\rm SLL}\right]_{11} &=&
 \begin{array}{c}
 -\f{3 N}2+\f5N+\f{2\lo}{N} +\lambda\left(\f1{2 N}+\f{2\lo}{N}\right),
 \end{array}\\
\nnb\left[\Delta r_{\overline{\rm MS}\rightarrow {\rm RI}}^{\rm SLL}\right]_{12} &=&
 \begin{array}{c}
 -\f72-2 \lo-\lambda\left(\f12+ 2\lo\right),
 \end{array}\\
\nnb\left[\Delta r_{\overline{\rm MS}\rightarrow {\rm RI}}^{\rm SLL}\right]_{13} &=&
 \begin{array}{c}
 \f{N}2-\f{13}{12 N}+\f{5\lo}{6 N}+\lambda\left(\f{N}8-\f1{6 N}+\f\lo{6 N}\right),
 \end{array}\\
\nnb\left[\Delta r_{\overline{\rm MS}\rightarrow {\rm RI}}^{\rm SLL}\right]_{14} &=&
 \begin{array}{c}
 \f{7}{12}-\f{5 \lo}{6} +\lambda\left(\f1{24}-\f{\lo}{6}\right),
 \end{array}\\
\nnb\left[\Delta r_{\overline{\rm MS}\rightarrow {\rm RI}}^{\rm SLL}\right]_{21} &=&
 \begin{array}{c}
 -1-2\lo +\lambda\left(1-2\lo\right),
 \end{array}\\
\nnb\left[\Delta r_{\overline{\rm MS}\rightarrow {\rm RI}}^{\rm SLL}\right]_{22} &=&
 \begin{array}{c}
 -4 N+\f5N+\f{2 \lo}{N}-\lambda\left(\f{3 N}2- \f1{2 N}-\f{2\lo}{N}\right),
 \end{array}\\
\nnb\left[\Delta r_{\overline{\rm MS}\rightarrow {\rm RI}}^{\rm SLL}\right]_{23} &=&
 \begin{array}{c}
 \f{13}{12}-\f{5\lo}6+\lambda\left(\f16-\f\lo6\right),
 \end{array}\\
\nnb\left[\Delta r_{\overline{\rm MS}\rightarrow {\rm RI}}^{\rm SLL}\right]_{24} &=&
 \begin{array}{c}
 -\f{13}{12 N}+\f{5 \lo}{6 N} -\lambda\left(\f1{6 N}-\f{\lo}{6 N}\right),
 \end{array}\\
\nnb\left[\Delta r_{\overline{\rm MS}\rightarrow {\rm RI}}^{\rm SLL}\right]_{31} &=&
 \begin{array}{c}
 4 N-\f{12}N +\f{40 \lo}N+\lambda\left(6 N-\f8N+\f{8\lo}{N}\right),
 \end{array}\\
\nnb\left[\Delta r_{\overline{\rm MS}\rightarrow {\rm RI}}^{\rm SLL}\right]_{32} &=&
 \begin{array}{c}
 8-40 \lo+\lambda\left(2-8 \lo\right),
 \end{array}\\
\nnb\left[\Delta r_{\overline{\rm MS}\rightarrow {\rm RI}}^{\rm SLL}\right]_{33} &=&
 \begin{array}{c}
 -\f{5 N}2-\f5{3 N}+\f{26\lo}{3 N}-\lambda\left(N+\f{5}{6 N}-\f{10 \lo}{3 N}\right),
 \end{array}\\
\nnb\left[\Delta r_{\overline{\rm MS}\rightarrow {\rm RI}}^{\rm SLL}\right]_{34} &=&
 \begin{array}{c}
 \f{25}{6}-\f{26 \lo}{3} +\lambda\left(\f{11}6-\f{10\lo}{3}\right),
 \end{array}\\
\nnb\left[\Delta r_{\overline{\rm MS}\rightarrow {\rm RI}}^{\rm SLL}\right]_{41} &=&
 \begin{array}{c}
 12 -40 \lo+\lambda\left(8-8\lo\right),
 \end{array}\\
\nnb\left[\Delta r_{\overline{\rm MS}\rightarrow {\rm RI}}^{\rm SLL}\right]_{42} &=&
 \begin{array}{c}
 -\f{12}N+\f{40 \lo}N-\lambda\left(\f8N-\f{8 \lo}{N}\right),
 \end{array}\\
\nnb\left[\Delta r_{\overline{\rm MS}\rightarrow {\rm RI}}^{\rm SLL}\right]_{43} &=&
 \begin{array}{c}
 \f53-\f{26\lo}3+\lambda\left(\f13-\f{10 \lo}3\right),
 \end{array}\\
\nnb\left[\Delta r_{\overline{\rm MS}\rightarrow {\rm RI}}^{\rm SLL}\right]_{44} &=&
 \begin{array}{c}
 -\f5{3 N}+\f{26 \lo}{3 N} +\lambda\left(\f{N}2-\f5{6 N}+\f{10\lo}{3 N}\right).
\end{array}
\nnb
\eea
\mathindent1cm
In section~\ref{sec:comp}, the above results will be used in
performing the comparison with ref.~\cite{CFLMSS97}.

\newsection{Recovering the ADM of $\Delta F=2$ operators\\ from 
$\Delta F=1$ results}
\label{sec:transf}

Let us now use our $\Delta F =1$ anomalous dimensions from
section~\ref{sec:df1} to find again the ADM of $\Delta F=2$ operators.
This will serve as a cross-check of our findings and as a preparation
for the comparison with ref.~\cite{CFLMSS97} in
section~\ref{sec:comp}.

Starting from eq.~(\ref{normal.df1}), we shall pass to another
operator basis where the operators are either symmetric or
antisymmetric under $d \leftrightarrow c$ interchange.  Next, the
flavours of both quarks and both antiquarks will be set equal.  For
definiteness, we shall do it first in the SLL sector.  The superscript
``SLL'' will be understood for all the relevant quantities below, and
we shall not write it explicitly.

In four spacetime dimensions, passing to the new operator basis would
be equivalent to performing a simple linear transformation of the
operators. In the framework of dimensional regularization, introducing
additional evanescent operators becomes necessary. In the SLL sector,
only two evanescent operators were needed in the $\Delta F = 1$
calculation (see appendix B). Now, we need to introduce six additional
evanescent operators in this sector. They are defined in appendix C.

We begin with a redefinition of the physical operators $Q_i~(i=1, ...,
4)$ that amounts to adding to them appropriate linear combinations of
the evanescent operators $E_i$:
\be \label{add.evan}
Q_i \to Q_i + \sum_{k=1}^8 W_{ik} E_k ~\equiv~ \left[Q_i\right]_{\rm new}
\ee
where
\be
\hat{W} = \left( \begin{array}{cccccccc}
0& 0& -\f{1}{2}& 0& \f{1}{8}& 0& 0& 0\\
0& 0&         0& 0&        0& 0& 0& 0\\
0& 0&         6& 0& \f{1}{2}& 0& 0& 0\\
0& 0&         0& 0&        0& 0& 0& 0
\end{array} \right).
\ee
The ``new'' operators read
\bea
\left[Q_1\right]_{\rm new} &=& -\frac12 \left[Q_1\right]_F
                               +\frac18 \left[Q_3\right]_F,
\hspace{2cm}
\left[Q_2\right]_{\rm new} = Q_2,\\
\left[Q_3\right]_{\rm new} &=& 6 \left[Q_1\right]_F + 
                               \frac12 \left[Q_3\right]_F,
\hspace{24mm}
\left[Q_4\right]_{\rm new} = Q_4,
\eea
where 
\be
\left[Q_1\right]_F=\left(\bar s^\alpha P_L c^\alpha\right)
                    \left(\bar u^\beta P_L d^\beta\right),
\qquad
\left[Q_3\right]_F=\left(\bar s^\alpha \sigma_{\mu\nu} P_L c^\alpha\right)
                   \left(\bar u^\beta \sigma^{\mu\nu} P_L d^\beta\right).
\ee
In 4 spacetime dimensions, the transformation (\ref{add.evan}) would
be equivalent to performing the Fierz rearrangement of $Q_1$ and
$Q_3$, as $E_k$ would not contribute.  Since the Fierz identities
cannot be analytically continued to $D$ dimensions, the Fierz
rearrangement {\em must} be understood in terms of the transformation
(\ref{add.evan}), so long as the $\overline{\rm MS}$ scheme is used.
The $\overline{\rm MS}$-renormalized one-loop matrix elements of $Q_1$
and $Q_3$ are affected by this transformation. This means that the
renormalization scheme is changed.  We pass from one version of the
NDR--$\overline{\rm MS}$ scheme to another, even though the evanescent
operators remain unchanged.

After the redefinition (\ref{add.evan}), we perform a simple linear
transformation of the operators
\be
\left[Q_i\right]_{\rm new} \to \sum_{j=1}^4 R_{ij} \left[Q_j\right]_{\rm new}
\ee
with
\be \label{RSLL}
\hat{R} = \left( \begin{array}{cccc}
-\f{1}{4}& \f{1}{2}&  \f{1}{16}&        0 \\[2mm]
        3&        0&   \f{1}{4}& \f{1}{2} \\[2mm]
 \f{1}{4}& \f{1}{2}& -\f{1}{16}&        0 \\[2mm]
       -3&        0&  -\f{1}{4}& \f{1}{2} 
\end{array} \right).
\ee
As one can easily check, our final operator basis is
$\{Q_1^+,Q_2^+,Q_1^-,Q_2^-\}$, where
\bea
Q_1^{\pm} &=& \f{1}{2} \left[ 
      (\bar{s}^{\alpha}                 P_L d^{\alpha})
      (\bar{u}^{ \beta}                 P_L c^{ \beta})
  \pm (\bar{s}^{\alpha}                 P_L c^{\alpha})
      (\bar{u}^{ \beta}                 P_L d^{ \beta}) \right],
\nnb \\[2mm]
Q_2^{\pm} &=& \f{1}{2} \left[ 
      (\bar{s}^{\alpha} \sigma_{\mu\nu} P_L d^{\alpha})
      (\bar{u}^{ \beta} \sigma^{\mu\nu} P_L c^{ \beta})
  \pm (\bar{s}^{\alpha} \sigma_{\mu\nu} P_L c^{\alpha})
      (\bar{u}^{ \beta} \sigma^{\mu\nu} P_L d^{ \beta}) \right].
\label{opm.SLL}
\eea
The ADM transforms as follows:
\bea
\hat{\gamma}^{(0)} &\to& \hat{R} \hat{\gamma}^{(0)} \hat{R}^{-1},
\label{transf0}\\
\hat{\gamma}^{(1)} &\to& \hat{R} \left\{ \hat{\gamma}^{(1)} 
+ \left[ \Delta \hat{r}, \hat{\gamma}^{(0)} \right]
        + 2 \beta_0 \Delta \hat{r} \right\} \hat{R}^{-1},
\label{transf1}
\eea
where $\beta_0 = \f{11}{3}N-\f{2}{3} f$. The matrix $\Delta \hat{r}$
reflects in the usual manner \cite{BJLW92} change of the
renormalization scheme that follows from eq.~(\ref{add.evan}). The
explicit form of $\Delta \hat{r}$ is \cite{CMM98.df}
\be
\Delta \hat{r} = -\hat{W} \hat{c},
\ee
provided $\hat{W} \hat{e}=0$. The matrices $\hat{c}$ and $\hat{e}$ are
found from one-loop matrix elements of evanescent operators, as in
eq.~(\ref{evan.melem}). The product $\hat{W} \hat{e}$ indeed vanishes
in our case, and
\be
\hat{c} = \left( \begin{array}{cccc}
\star & \star & \star & \star \\
\star & \star & \star & \star \\
\f{3}{4} N -\f{5}{N}  & \f{17}{4} & ~~~\f{1}{16} N -\f{1}{4 N}~~~ &  \f{3}{16} \\ 
\star & \star & \star & \star \\ ~~~
7 N-\f{28}{N}~~~ &        21   &      -\f{7}{4} N +\f{5}{N}    & -\f{13}{4} \\
\star & \star & \star & \star \\
\star & \star & \star & \star \\
\star & \star & \star & \star 
\end{array} \right).
\ee
Here, stars denote non-vanishing elements of $\hat{c}$ that are
irrelevant for us, since they do not affect the matrix 
\be
\Delta\hat{r}=-\hat{W} \hat{c} =
\left( \begin{array}{cccc}
-\f{1}{2} N + \f{1}{N} & ~~-\f{1}{2}~~ & ~~\f{1}{4}N -\f{3}{4 N}~~ & \f{1}{2} \\[2mm] 
0 & 0 & 0 & 0 \\[2mm]
-8N + \f{44}{N} & -36 & \f{1}{2} N -\f{1}{N} & \f{1}{2} \\[2mm]
0 & 0 & 0 & 0 
\end{array} \right).
\ee
After the transformation (\ref{transf0}, \ref{transf1}), the ADM in the
basis $\{Q_1^+,Q_2^+,Q_1^-,Q_2^-\}$ is found to have the form
\be
\hat{\gamma}_{4\times4} = \left( \begin{array}{cc}
\hat{\gamma}_{2\times2}^+ & 0_{2\times2} \\
0_{2\times2} & \hat{\gamma}_{2\times2}^- \end{array} \right),
\ee
where $\hat{\gamma}^{\pm} = \hat{\gamma}^{(0)\pm} + {\displaystyle
  \f{\al}{4 \pi}} \hat{\gamma}^{(1)\pm} +...$,
\be
\hat{\gamma}^{(0)\pm} = \left( \begin{array}{ccc}
-6 N \pm 6 +\f{6}{N} &~& \pm \f{1}{2} -\f{1}{N} \\[1mm] 
\mp 24 -\f{48}{N}    &~& 2 N \pm 6 -\f{2}{N} 
\end{array} \right), \\[2mm]
\ee
and
\be  \label{gamma.pm.SLL}
\begin{array}{rcl}
 \gamma^{(1)\pm}_{11} &=& -\f{203}{6} N^2  \pm \f{107}{3} N +\f{136}{3}
\mp \f{12}{N} -\f{107}{2 N^2} +\f{10}{3} N f \mp \f{2}{3} f -\f{10}{3 N} f,
\\[2mm]
\gamma^{(1)\pm}_{12} &=& \mp \f{1}{36} N -\f{31}{9} \pm \f{9}{N} -\f{4}{N^2}
 \mp \f{1}{18} f +\f{1}{9 N} f,
\\[2mm]
\gamma^{(1)\pm}_{21} &=& \mp \f{364}{3} N -\f{704}{3} \mp \f{208}{N}
-\f{320}{N^2} \pm \f{136}{3} f + \f{176}{3 N} f,
\\[2mm]
\gamma^{(1)\pm}_{22} &=& \f{343}{18} N^2 \pm 21 N -\f{188}{9} \pm \f{44}{N}
+\f{21}{2 N^2} -\f{26}{9} N f \mp 6 f +\f{2}{9 N} f.
\end{array}
\ee

One can easily verify that the matrix $\hat{\gamma}^+$ is equal to the
one we have already found in eqs.~(\ref{g0df2}) and (\ref{g1df2}). It
must be so, because the operators $Q_i^+$ from eq.~(\ref{opm.SLL})
reduce to $Q_i^{\rm SLL}$ from eq.~(\ref{normal}) when the flavour
replacements $c\to d$ and $\bar{u}\to\bar{s}$ are made.  Moreover, the
evanescent operators listed in appendices B and C can be linearly
combined to the ones that are either symmetric or antisymmetric under
$d \leftrightarrow c$ interchange. When the flavour replacements $c\to
d$ and $\bar{u}\to\bar{s}$ are made, the antisymmetric operators
vanish, while the symmetric ones become equal to those in appendix A.
Thus, we have shown how to extract the $\Delta F=2$ results from
the $\Delta F=1$ ones.

Let us now briefly describe the analogous transformations in the VLL
and LR$\equiv$VLR$\oplus$SLR sectors. All the necessary evanescent
operators are given in appendices B and C. The relevant matrices
$\hat{W}$ and $\hat{R}$ are the following:
\bea
\hat{W}^{\rm VLL} &=& \left( \begin{array}{cccccc}
0 & 0 & 1 & 0 & 0 & 0 \\
0 & 0 & 0 & 0 & 0 & 0 \end{array} \right),
\nnb\\[2mm]
\hat{R}^{\rm VLL} &=& \f{1}{2} \left( \begin{array}{cc}
 1 & 1 \\
-1 & 1 \end{array} \right),
\nnb\\[2mm]
\hat{W}^{\rm LR}_{4 \times 12} &\equiv& \left( \begin{array}{cc}
\hat{W}^{\rm VLR}_{2 \times 6} & 0_{2 \times 6} \\ 
0_{2 \times 6} & \hat{W}^{\rm SLR}_{2 \times 6} 
\end{array} \right) =
\left( \begin{array}{cc}
-2 \hat{W}^{\rm VLL} & 0_{2 \times 6} \\
0_{2 \times 6} & -\f{1}{2} \hat{W}^{\rm VLL}
\end{array} \right),
\nnb\\[2mm]
\hat{R}^{\rm LR} &=& \f{1}{2} \left( \begin{array}{cccc}
        0 & 1 & -2 & 0 \\
-\f{1}{2} & 0 &  0 & 1 \\
        0 & 1 &  2 & 0 \\
 \f{1}{2} & 0 &  0 & 1
\end{array} \right).
\eea
Consequently, the final operator bases are
$\{Q_1^{VLL+},Q_1^{VLL-}\}$ and
$\{Q_1^{LR+},Q_2^{LR+},Q_1^{LR-},Q_2^{LR-}\}$, where
\bea
Q_1^{VLL\pm} &=& \f{1}{2} \left[ 
      (\bar{s}^{\alpha} \gamma_{\mu} P_L d^{\alpha})
      (\bar{u}^{ \beta} \gamma^{\mu} P_L c^{ \beta})
  \pm (\bar{s}^{\alpha} \gamma_{\mu} P_L c^{\alpha})
      (\bar{u}^{ \beta} \gamma^{\mu} P_L d^{ \beta}) \right],
\nnb \\[2mm]
Q_1^{LR\pm} &=& \f{1}{2} \left[ 
      (\bar{s}^{\alpha} \gamma_{\mu} P_L d^{\alpha})
      (\bar{u}^{ \beta} \gamma^{\mu} P_R c^{ \beta})
  \pm (\bar{s}^{\alpha} \gamma_{\mu} P_R c^{\alpha})
      (\bar{u}^{ \beta} \gamma^{\mu} P_L d^{ \beta}) \right],
\nnb \\[2mm]
Q_2^{LR\pm} &=& \f{1}{2} \left[ 
      (\bar{s}^{\alpha}              P_L d^{\alpha})
      (\bar{u}^{ \beta}              P_R c^{ \beta})
  \pm (\bar{s}^{\alpha}              P_R c^{\alpha})
      (\bar{u}^{ \beta}              P_L d^{ \beta}) \right].
\label{opm.VLL.LR}
\eea

An important simplification in the present case is that the one-loop
matrix elements of the evanescent operators $E_3^{\rm VLL}$, $E_3^{\rm VLR}$
and $E_3^{\rm SLR}$ from appendix C vanish in the limit $D\to4$, after
subtraction of the MS-counterterms proportional to evanescent
operators only.  This means that the third rows of $\hat{c}^{\rm VLL}$,
$\hat{c}^{\rm VLR}$ and $\hat{c}^{\rm SLR}$ vanish (cf.
eq.~(\ref{evan.melem})). Consequently, $\Delta \hat{r}^{\rm VLL} =
-\hat{W}^{\rm VLL} \hat{c}^{\rm VLL}=0$ and $\Delta \hat{r}^{\rm LR} =
-\hat{W}^{\rm LR} \hat{c}^{\rm LR}=0$. This is why the two-loop \linebreak
$\Delta F=1$ matrices of the VLL, VLR and SLR sectors exhibited Fierz
symmetry in eq.~(\ref{fierz.rel}). The transformations of the two-loop
ADMs in the VLL and LR sectors thus look as if we worked in 4
dimensions, i.e.  they reduce to simple multiplications by the
corresponding $\hat{R}$-matrices and their inversions. The final
results are
\be 
\hat{\gamma}^{\rm VLL}_{2\times 2} = \left( \begin{array}{cc}
    \hat{\gamma}^{VLL+} & 0 \\
    0 & \hat{\gamma}^{VLL-} \end{array} \right),
\hspace{3cm}
\hat{\gamma}^{\rm LR}_{4\times 4} = \left( \begin{array}{cc}
    \hat{\gamma}_{2\times 2}^{LR+} & 0_{2\times 2} \\[2mm]
    0_{2\times 2} & \hat{\gamma}_{2\times 2}^{LR-} \end{array} \right),
\ee
where
\bea
\gamma^{(0)VLL\pm} &=& \pm 6 - \f{6}{N}, 
\nnb\\[2mm]
\gamma^{(1)VLL\pm} &=& \mp \f{19}{6} N -\f{22}{3} \pm \f{39}{N} -\f{57}{2 N^2} 
                 \pm \f{2}{3} f -\f{2}{3 N} f,
\nnb\\[2mm]
\hat{\gamma}^{(0)LR\pm} &=& \left( \begin{array}{ccc} 
\f{6}{N} && \pm 12 \\[1mm] 
0 && - 6 N + \f{6}{N} \end{array} \right),
\nnb\\[2mm]
\hat{\gamma}^{(1)LR\pm} &=& \left( \begin{array}{ccc} 
\f{137}{6} + \f{15}{2 N^2} - \f{22}{3N} f 
&~~~&
\pm \f{200}{3} N \mp \f{6}{N} \mp \f{44}{3} f 
\\[1mm] 
\pm \f{71}{4} N \pm \f{9}{N} \mp 2f 
&&
-\f{203}{6} N^2 +\f{479}{6} +\f{15}{2 N^2} +\f{10}{3} N f -\f{22}{3N} f 
\end{array} \right).
\eea
One can see that $\gamma^{VLL+}$ and $\hat{\gamma}^{LR+}$ are
identical to our $\Delta F=2$ results in eqs.~(\ref{gVLL.df2}),
(\ref{g0LR.df2}) and (\ref{g1LR.df2}).

\newpage
\newsection{Comparison with previous ADM calculations}
\label{sec:comp}

In the present section, we compare our findings from
sections~\ref{sec:df2}, \ref{sec:df1} and \ref{sec:penguin} with
the previously published results for anomalous-dimension matrices.

\subsection{One-loop results}

As far as the one-loop QCD ADMs of four-quark operators are
concerned, the historical order of their evaluation was as follows:
\begin{itemize} 
\item{} Current--current contributions to the one-loop ADM of $\Delta
  F=1$ operators belonging to the VLL and VLR sectors were originally
  calculated in refs.~\cite{AM74,GL74}. These results were also
  immediately applicable to the SLR sector, because the Fierz
  rearrangement has a trivial effect at one loop. For the same reason,
  one-loop anomalous dimensions of the $\Delta F=2$ operators
  belonging to the VLL and LR sectors could have been immediately read
  off from these articles. Thus, after 1974, the only unpublished
  one-loop current--current anomalous dimensions were those of the SLL
  sector, both in the $\Delta F=1$ and $\Delta F=2$ cases.
\item{} One-loop penguin contributions to the ADM of the Standard
  Model operators were originally evaluated in
  refs.~\cite{VZS77}--\cite{GP80}.  As we have shown in
  section~\ref{sec:penguin}, penguin contributions to the ADM of other
  (beyond--SM) flavour-changing dimension-six operators can be easily
  extracted from the SM calculations, both at one and at two loops.
\item{} To our knowledge, the first published results for
  $\gamma^{(0)\rm SLL}$ occur in refs.~\cite{BMZ91} and \cite{CFLMSS97},
  for the $\Delta F=2$ and $\Delta F=1$ cases, respectively.
\end{itemize}
The one-loop ADMs given in the present article agree with all the
papers quoted above. However, in order to perform comparisons, one
often needs to make simple linear transformations, because different
operator bases are used by different authors. For instance, the
results for $\hat{\gamma}^{(0)\rm SLL}$ in ref.~\cite{BMZ91} are given in
the basis $\{Q_1^{\rm SLL} ,\tilde{Q}_1^{\rm SLL}\}$. In order to compare them
with our eq.~(\ref{g0df2}), one should use the relation
(\ref{opfSLL1}). Similarly, eqs.~(\ref{RSLL}) and (\ref{transf0}) need
to be used for comparing our $\hat{\gamma}^{(0)\rm SLL}$~ in
eq.~(\ref{ga0SLL}) with the corresponding results in
ref.~\cite{CFLMSS97}.

\subsection{Two-loop results}

The history of previous two-loop computations is as follows:
\begin{itemize}
\item{} The current--current anomalous dimensions of the $\Delta F=1$
  operators belonging to the VLL sector were originally calculated in
  ref.~\cite{ACMP81} (in the DRED--$\overline{\rm MS}$ scheme), and
  confirmed in ref.~\cite{BW90} (where the NDR--$\overline{\rm MS}$
  and HV--$\overline{\rm MS}$ results were also given).
\item{} The remaining elements of the two-loop QCD ADM for $\Delta
  F=1$ operators relevant in the SM were calculated in
  refs.~\cite{BJLW93,CFMR94}. New results in these papers were the
  current--current contributions in the VLR sector, as well as all the
  penguin contributions. The SLR sector results in the $\Delta F=1$
  case, as well as the $\Delta F=2$ results for the VLL and LR sectors
  could be easily derived from them with the help of Fierz identities,
  because the NDR--$\overline{\rm MS}$-renormalized one-loop matrix
  elements remain invariant under Fierz transformations, except for
  the current--current ones in the SLL sector, and the penguin ones in
  the VLL sector.  Therefore, in the early 1990's, the only unknown
  two-loop anomalous dimensions were those of the SLL sector.
\item{} The first calculation of the two-loop ADM in the SLL sector
  was performed by Ciuchini et al.~\cite{CFLMSS97}, in both the
  $\Delta F=1$ and $\Delta F=2$ cases. The ADM was calculated there in
  the so-called ``FRI'' renormalization scheme.  The transformation rules
  were given to the LRI scheme (Landau-gauge RI-scheme) and to the
  NDR--$\overline{\rm MS}$ scheme. Current--current anomalous
  dimensions for the remaining sectors were recalculated as well.
\item{} Penguin contributions to the ADM of non-SM operators are
  considered for the first time in the present article.
\end{itemize}
All the two-loop results presented here agree with the previous
calculations mentioned above, except for the NDR--$\overline{\rm MS}$
ones for the SLL sector found in ref.~\cite{CFLMSS97}. Below, we
explain the reason for this disagreement.

\subsection{Comparison with ref.~~\cite{CFLMSS97}}

In ref.~\cite{CFLMSS97}, the two-loop ADM for $\Delta F=1$ operators
of the SLL sector was given in the basis defined in eq.~(13) of that
paper, which is equivalent to our eq.~(\ref{opm.SLL}). It was
presented in the so-called ``FRI'' renormalization scheme, and the
transformation rules to the NDR--$\overline{\rm MS}$ scheme were
appended. Applying these transformation rules to their ``FRI''-scheme
ADM, one obtains results that differ from our
eq.~(\ref{gamma.pm.SLL}). In particular, a mixing between $Q_i^-$ and
$Q_i^+$ occurs, which is absent in our result (\ref{gamma.pm.SLL}).
We could obtain their result if we ignored the transformation
(\ref{add.evan}) and, consequently, used $\Delta \hat{r}=0$ in our
eq.~(\ref{transf1}). However, the final results would then correspond
to the basis
\mathindent0cm
\bea 
{Q'}_1^{\pm} &=& \f{1}{2} \left[ 
      (\bar{s}^{\alpha}                 P_L d^{\alpha})
      (\bar{u}^{ \beta}                 P_L c^{ \beta})
  \mp \f{1}{2} 
      (\bar{s}^{\alpha}                 P_L d^{\beta })
      (\bar{u}^{\beta }                 P_L c^{\alpha}) 
  \pm \f{1}{8}
      (\bar{s}^{\alpha} \sigma_{\mu\nu} P_L d^{ \beta})
      (\bar{u}^{ \beta} \sigma^{\mu\nu} P_L c^{\alpha}) \right],
\nnb \\[2mm]
{Q'}_2^{\pm} &=& \f{1}{2} \left[ 
      (\bar{s}^{\alpha} \sigma_{\mu\nu} P_L d^{\alpha})
      (\bar{u}^{ \beta} \sigma^{\mu\nu} P_L c^{ \beta})
\pm 6 (\bar{s}^{\alpha}                 P_L d^{\beta })
      (\bar{u}^{\beta }                 P_L c^{\alpha}) 
\pm \f{1}{2}
      (\bar{s}^{\alpha} \sigma_{\mu\nu} P_L d^{ \beta})
      (\bar{u}^{ \beta} \sigma^{\mu\nu} P_L c^{\alpha}) \right],\nnb\\
\label{unfierzed}
\eea
\mathindent1cm
rather than the one in eq.~(\ref{opm.SLL}). In 4 spacetime dimensions,
the operators (\ref{opm.SLL}) and (\ref{unfierzed}) are identical,
thanks to the Fierz identities (\ref{fSLL}).  However, in $D$
dimensions they are not. Consequently, their NDR--$\overline{\rm
  MS}$-renormalized matrix elements differ at one loop, and it
is not surprising that the two-loop ADM depends on which of the two
bases is used.

We informed the authors of ref.~\cite{CFLMSS97} about our findings
prior to publication of the present article. They responded that
although their NDR--$\overline{\rm MS}$ results had been claimed to
correspond to the basis (\ref{opm.SLL}), the NDR--$\overline{\rm MS}$
renormalization conditions had been actually imposed in the basis
(\ref{unfierzed}). However, they had forgotten to mention this in
their article. Unfortunately, such a mistake in the presentation has
the same effect on the final result as a mistake in the calculation
that amounts to missing $\Delta \hat{r} \neq 0$ in eq.~(\ref{transf1}).

As far as the two-loop ADM for $\Delta F=2$ operators of the SLL
sector is concerned, the situation is as follows. If we made the
flavour replacements $c\to d$ and $\bar{u}\to\bar{s}$ in the basis
(\ref{unfierzed}), but did not change anything in the ADM, we could
interpret this ADM as the one for $\Delta F=2$ operators, as the
authors of ref.~\cite{CFLMSS97} did. However, it would correspond to
quite non-standard conventions for the treatment of the evanescent
operators obtained from ${Q'}_1^{-}$ and ${Q'}_2^{-}$ after the
flavour replacements. One would need to assume that the finite
one-loop matrix elements of these evanescent operators are not
renormalized away, contrary to the usual procedure for any evanescent
operator \cite{BW90,CMM98.df,DG91,HN95}. Such non-standard conventions
make the RGE evolution more complicated, because one has to deal with
a $4\times 4$ instead of a $2 \times 2$~ ADM in the
NDR--$\overline{\rm MS}$ RGE for the SLL sector, in the $\Delta F=2$
case. The calculation of the one-loop matrix elements becomes more
involved, as well.

In the $\Delta F=2$ case, no calculation is necessary to convince
oneself that the results of ref.~\cite{CFLMSS97} cannot correspond to
the NDR--$\overline{\rm MS}$ renormalization conditions imposed in the
basis (\ref{opm.SLL}) (their eq.~(13)).  Once the $c\to d$ and
$\bar{u}\to\bar{s}$ replacements have been made, the operators $Q_i^-$
in eq.~(\ref{opm.SLL}) vanish identically in $D$ dimensions.
Therefore, they cannot mix into the $Q_i^+$ operators, independently
of what the treatment of evanescent operators is.  On the other hand,
mixing of $Q_i^-$ into $Q_i^+$ was claimed to be found in the
NDR--$\overline{\rm MS}$ scheme in ref.~\cite{CFLMSS97}.  Therefore,
an inconsistency is clearly seen.

In the remainder of this section, we shall verify that our
NDR--$\overline{\rm MS}$ results are compatible with the LRI ones of
ref.~\cite{CFLMSS97}. By differentiating eq.~(\ref{coeff.transf}) with
respect to $\mu$, one obtains
\bea
\hat{\gamma}^T_{\rm RI}(\mu) \vec{C}^{\rm RI}(\mu) &=& \left[
\f{\beta_0 \al^2(\mu)}{8\pi^2} \Delta \hat{r}^T_{\overline{\rm MS}\to RI}(\mu) 
+  \f{\beta^0_{\lambda} \al(\mu)}{8\pi^2} \lambda(\mu) 
\left( \f{\partial}{\partial\lambda} 
         \Delta \hat{r}^T_{\overline{\rm MS}\to RI}(\mu) \right) 
\right. \nnb\\ && \left. +\left( 1 - 
\f{\al(\mu)}{4\pi} \Delta \hat{r}^T_{\overline{\rm MS}\to RI}(\mu) \right) 
\hat{\gamma}^T_{\overline{\rm MS}}(\mu) \right] 
\vec{C}^{\overline{\rm MS}}(\mu) ~+~ {\cal O}\left(\al^3\right),
\label{diff.c}
\eea
where we have used the RGE (\ref{RGE}),
\be
\mu \f{d}{d\mu} \al(\mu) = -\f{\beta_0 \al(\mu)^2}{2\pi} + {\cal O}(\al^3)
\hspace{8mm} {\rm and} \hspace{8mm}
\mu \f{d}{d\mu} \lambda(\mu) = 
-\f{\beta^0_{\lambda} \al(\mu)}{2\pi} \lambda(\mu) + {\cal O}(\al^2).
\hspace{8mm}
\ee
We have also used the fact that the dependence of
$\Delta \hat{r}_{\overline{\rm MS}\to RI}$ on $\mu$ originates solely from
its dependence on the gauge-fixing parameter $\lambda(\mu)$.

Next, we use eq.~(\ref{coeff.transf}) again to express
$\vec{C}^{\overline{\rm MS}}(\mu)$ by $\vec{C}^{\rm RI}(\mu)$ in
eq.~(\ref{diff.c}). Then, the first two terms of the perturbative
expansion (\ref{gamma.exp}) of $\hat{\gamma}_{\rm RI}$ can be easily
read off 
\bea
\hat{\gamma}^{(0)}_{\rm RI} &=& \hat{\gamma}^{(0)}_{\overline{\rm MS}},\\
\hat{\gamma}^{(1)}_{\rm RI} &=& \hat{\gamma}^{(1)}_{\overline{\rm MS}}
+ \left[ \Delta \hat{r}_{\overline{\rm MS}\to RI},~
         \hat{\gamma}^{(0)}_{\overline{\rm MS}} \right]
+ 2 \left( \beta_0 + \beta^0_{\lambda}\;\lambda \f{\partial}{\partial\lambda} \right)
\Delta \hat{r}_{\overline{\rm MS}\to RI}.
\eea

Armed with our explicit expressions for $\Delta \hat{r}_{\overline{\rm
    MS}\to RI}$ given in section~\ref{sec:RI} and with the values of
\be
\beta_0 = \f{11}{3} N - \f{2}{3} f 
\hspace{2cm} {\rm and} \hspace{2cm}
\beta^0_{\lambda} = \left( \f{\lambda}{2} - \f{13}{6} \right) N + \f{2}{3} f,
\ee
we can easily calculate the RI-scheme ADM from our ${\overline{\rm
    MS}}$ results, for arbitrary $\lambda$.  Setting then $\lambda \to
0$, we recover all the LRI-scheme anomalous dimensions given in
ref.~\cite{CFLMSS97}.

As far as the ``FRI''-scheme ADMs of ref.~\cite{CFLMSS97} are
concerned, we can confirm them as well. However, it should be
emphasized that the ``FRI'' scheme is not equivalent to the RI scheme
considered in section \ref{sec:RI} for any choice of $\lambda$.  The
``FRI'' scheme cannot be defined beyond perturbation theory, because
different external momenta are chosen in different diagrams when the
renormalization conditions are specified.  Therefore, in our opinion,
the main advantage of the RI scheme is lost.

\newsection{Conclusions}
\label{sec:concl}

In the present paper, we have calculated the two-loop QCD anomalous
dimensions matrix (ADM) ~$(\hat \gamma^{(1)})_{NDR}$~ in the
NDR--$\overline{\rm MS}$ scheme for all the four-fermion dimension-six
flavour-changing operators that are relevant to both the Standard Model
and its extensions. 

The $\Delta F=2$ two-loop results can be found in eqs.~(\ref{g1df2}),
(\ref{gVLL.df2}) and (\ref{g1LR.df2}). While the matrices in
eqs.~(\ref{gVLL.df2}) and (\ref{g1LR.df2}) could be extracted from the
already published results, the two-loop NDR--$\overline{\rm MS}$ ADM
(\ref{g1df2}) for the SLL operators defined in eq.~(\ref{normal}) is
correctly calculated for the first time here.

The $\Delta F=1$ two-loop results for operators containing four
different quark flavours can be found in eqs.~(\ref{ga1VLL}),
(\ref{ga1VLR}), (\ref{ga1SLR}) and (\ref{ga1SLL}). While the matrices
in eqs.~(\ref{ga1VLL}), (\ref{ga1VLR}) and (\ref{ga1SLR}) could be
extracted from the already published results, the two-loop
NDR--$\overline{\rm MS}$ ADM (\ref{ga1SLL}) for the SLL operators
defined in eq.~(\ref{normal.df1}) is correctly calculated for the
first time here.

Penguin contributions to the ADM of non-SM operators have been
considered for the first time here. These contributions can be easily
extracted from the existing SM calculations. We have identified the
relevant non-SM operators in the $\Delta S=1$ case, and presented the
corresponding ADM explicitly in eqs.~(\ref{ga0peng}) and
(\ref{ga1peng}).

We have demonstrated that the main findings of our paper, given in
eqs.~(\ref{g1df2}) and (\ref{ga1SLL}), are compatible with each other,
i.e. we have shown how to properly transform the ADMs from the
$\Delta F=1$ to the $\Delta F=2$ case. In this context, we have pointed
out that in the process of this transformation it is necessary to
introduce additional evanescent operators that vanish in four
spacetime dimensions because of the Fierz identities.

We have also given the rules that allow transforming our
NDR--$\overline{\rm MS}$ ADMs to the corresponding results in the RI
scheme, for arbitrary gauge-fixing parameter $\lambda$. They can be
found in the end of section 5. 

The $\Delta F=1$ two-loop ADMs for all the operators defined in
eq.~(\ref{normal.df1}) were previously presented in
ref.~\cite{CFLMSS97}, in the $Q_i^{\pm}$ basis. In the case of VLL,
VLR and SLR operators, there is full agreement between their and our
results. The case of SLL operators is more subtle. We can confirm
their LRI-scheme results (RI scheme with $\lambda=0$). However, their
NDR--$\overline{\rm MS}$ ADM is compatible with ours {\em only} after
correcting their eq.~(13), i.e. after changing the definitions of
their SLL operators to the ones given in eq.~(\ref{unfierzed}).

After such a correction in eq.~(13) of ref.~\cite{CFLMSS97}, also
their $\Delta F=2$ NDR--$\overline{\rm MS}$ results are compatible with
ours, provided they are understood in terms of quite non-standard
conventions for the treatment of evanescent operators. In their
conventions, the two-loop $\Delta F=2$ NDR--$\overline{\rm MS}$ ADM is
a 4$\times$4 rather than 2$\times$2 matrix, which makes the RGE
evolution and calculating low-energy matrix elements unnecessarily
complicated. Consequently, the results presented here should be more
useful for phenomenological applications.

\ \\
{\Large\bf Acknowledgements}

We thank the authors of ref.~\cite{CFLMSS97} for extensive discussions
concerning their paper. Furthermore, we would like to thank Christoph
Bobeth and Gerhard Buchalla for carefully reading the manuscript.
A.B. and J.U. acknowledge support from the German Bundesministerium
f{\"u}r Bildung und Forschung under the contract O5HT9WOA0. M.M. has
been supported in part by the Polish Committee for Scientific Research
under grant 2~P03B~014~14, 1998-2000.

\ \\
{\Large\bf Appendix A}
\def\theequation{A.\arabic{equation}}

Here, we specify the evanescent operators that are necessary as
counterterms for one-loop diagrams with insertions of the $\Delta
F=2$ operators (\ref{normal}).
\bea
E_1^{\rm VLL} &=& 
(\bar{s}^{\alpha} \gamma_{\mu} P_L d^{ \beta})
(\bar{s}^{ \beta} \gamma^{\mu} P_L d^{\alpha})  - Q_1^{\rm VLL},
\nnb \\
E_2^{\rm VLL} &=& 
(\bar{s}^{\alpha} \gamma_{\mu} 
                  \gamma_{\nu} 
                  \gamma_{\rho} P_L d^{\alpha})
(\bar{s}^{ \beta} \gamma^{\mu} 
                  \gamma^{\nu} 
                  \gamma^{\rho} P_L d^{ \beta}) + (-16 + 4 \e) Q_1^{\rm VLL},
\nnb \\ 
E_3^{\rm VLL} &=& 
(\bar{s}^{\alpha} \gamma_{\mu} 
                  \gamma_{\nu} 
                  \gamma_{\rho} P_L d^{ \beta})
(\bar{s}^{ \beta} \gamma^{\mu} 
                  \gamma^{\nu} 
                  \gamma^{\rho} P_L d^{\alpha}) + (-16 + 4 \e) Q_1^{\rm VLL},
\nnb \\ 
E_1^{\rm LR} &=&  
(\bar{s}^{\alpha} P_L d^{ \beta})
(\bar{s}^{ \beta} P_R d^{\alpha}) + \f{1}{2} Q_1^{\rm LR},
\nnb \\ 
E_2^{\rm LR} &=&  
(\bar{s}^{\alpha} \gamma_{\mu}  P_L d^{ \beta})
(\bar{s}^{ \beta} \gamma^{\mu}  P_R d^{\alpha}) + 2 Q_2^{\rm LR},
\nnb \\ 
E_3^{\rm LR} &=& 
(\bar{s}^{\alpha} \gamma_{\mu} 
                  \gamma_{\nu} 
                  \gamma_{\rho} P_L d^{\alpha})
(\bar{s}^{ \beta} \gamma^{\mu} 
                  \gamma^{\nu} 
                  \gamma^{\rho} P_R d^{ \beta}) + (-4 - 4 \e) Q_1^{\rm LR},
\nnb \\ 
E_4^{\rm LR} &=& 
(\bar{s}^{\alpha} \gamma_{\mu} 
                  \gamma_{\nu} 
                  \gamma_{\rho} P_L d^{ \beta})
(\bar{s}^{ \beta} \gamma^{\mu} 
                  \gamma^{\nu} 
                  \gamma^{\rho} P_R d^{\alpha}) + (8 + 8 \e) Q_2^{\rm LR},
\nnb \\ 
E_5^{\rm LR} &=& 
(\bar{s}^{\alpha} \sigma_{\mu\nu} P_L d^{\alpha})
(\bar{s}^{ \beta} \sigma^{\mu\nu} P_R d^{ \beta}) - 6 \e Q_2^{\rm LR},
\nnb \\ 
E_6^{\rm LR} &=& 
(\bar{s}^{\alpha} \sigma_{\mu\nu} P_L d^{ \beta})
(\bar{s}^{ \beta} \sigma^{\mu\nu} P_R d^{\alpha}) + 3 \e Q_1^{\rm LR},
\nnb \\ 
E_1^{\rm SLL} &=&  
(\bar{s}^{\alpha} P_L d^{ \beta})
(\bar{s}^{ \beta} P_L d^{\alpha}) + \f{1}{2} Q_1^{\rm SLL} - \f{1}{8} Q_2^{\rm SLL},
\nnb \\ 
E_2^{\rm SLL} &=& 
(\bar{s}^{\alpha} \sigma_{\mu\nu} P_L d^{\beta})
(\bar{s}^{ \beta} \sigma^{\mu\nu} P_L d^{\alpha})
        - 6 Q_1^{\rm SLL} - \f{1}{2} Q_2^{\rm SLL},
\nnb \\ 
E_3^{\rm SLL} &=& 
(\bar{s}^{\alpha} \gamma_{\mu} 
                  \gamma_{\nu} 
                  \gamma_{\rho} 
                  \gamma_{\sigma} P_L d^{\alpha})
(\bar{s}^{ \beta} \gamma^{\mu} 
                  \gamma^{\nu} 
                  \gamma^{\rho} 
                  \gamma^{\sigma} P_L d^{ \beta})
              + (-64 + 96 \e) Q_1^{\rm SLL} + (-16 + 8 \e) Q_2^{\rm SLL},
\nnb \\ 
E_4^{\rm SLL} &=& 
(\bar{s}^{\alpha} \gamma_{\mu} 
                  \gamma_{\nu} 
                  \gamma_{\rho} 
                  \gamma_{\sigma} P_L d^{ \beta})
(\bar{s}^{ \beta} \gamma^{\mu} 
                  \gamma^{\nu} 
                  \gamma^{\rho} 
                  \gamma^{\sigma} P_L d^{\alpha})
              -64 Q_1^{\rm SLL} + (-16 + 16 \e) Q_2^{\rm SLL}.
\nnb
\eea

The evanescent operators for the VRR and SRR sectors, i.e.
$E_k^{VRR}$ and $E_k^{SRR}$ are obtained by replacing $L$ by $R$ in
the definitions of $E_k^{\rm VLL}$ and $E_k^{\rm SLL}$.

The operators $E_1^{\rm VLL}$, $E_1^{\rm LR}$, $E_2^{\rm LR}$, $E_1^{\rm SLL}$ and
$E_2^{\rm SLL}$ vanish in four spacetime dimensions because of the Fierz
identities (\ref{fVLL}), (\ref{fLR}) and (\ref{fSLL}). The operators
$E_2^{\rm VLL}$, $E_3^{\rm VLL}$, $E_3^{\rm LR}$, $E_4^{\rm LR}$, $E_3^{\rm SLL}$ and
$E_4^{\rm SLL}$ vanish by the four-dimensional identity
(\ref{three.gammas}).  Finally, $E_5^{\rm LR}$ and $E_6^{\rm LR}$ vanish in
four dimensions, because they become full contractions of self-dual
and self-antidual antisymmetric tensors.

The evanescent operators listed here would look somewhat simpler if we
removed from them all the terms proportional to $\e$. It would be
equivalent to changing one version of the $\overline{\rm MS}$ scheme
to another. Keeping the terms proportional to $\e$ in the above
equations makes our NDR--$\overline{\rm MS}$ scheme equivalent to the
one where the so-called ``Greek projections'' %\cite{B98,BW90,TV82} 
are used (see appendix D).

\ \\
{\Large\bf Appendix B}
\def\theequation{B.\arabic{equation}}

Here, we specify the evanescent operators that are necessary as
counterterms for one-loop diagrams with insertions of the $\Delta
F=1$ operators (\ref{normal.df1}).
\bea
E_1^{\rm VLL} &=& 
(\bar{s}^{\alpha} \gamma_{\mu} 
                  \gamma_{\nu} 
                  \gamma_{\rho} P_L d^{\beta})
(\bar{u}^{ \beta} \gamma^{\mu} 
                  \gamma^{\nu} 
                  \gamma^{\rho} P_L c^{\alpha}) + (-16 + 4 \e) Q_1^{\rm VLL},
\nnb \\
E_2^{\rm VLL} &=& 
(\bar{s}^{\alpha} \gamma_{\mu} 
                  \gamma_{\nu} 
                  \gamma_{\rho} P_L d^{\alpha})
(\bar{u}^{ \beta} \gamma^{\mu} 
                  \gamma^{\nu} 
                  \gamma^{\rho} P_L c^{ \beta}) + (-16 + 4 \e) Q_2^{\rm VLL},
\nnb \\ 
E_1^{\rm VLR} &=& 
(\bar{s}^{\alpha} \gamma_{\mu} 
                  \gamma_{\nu} 
                  \gamma_{\rho} P_L d^{\beta})
(\bar{u}^{ \beta} \gamma^{\mu} 
                  \gamma^{\nu} 
                  \gamma^{\rho} P_R c^{\alpha}) + (-4 - 4 \e) Q_1^{\rm VLR},
\nnb \\ 
E_2^{\rm VLR} &=& 
(\bar{s}^{\alpha} \gamma_{\mu} 
                  \gamma_{\nu} 
                  \gamma_{\rho} P_L d^{\alpha})
(\bar{u}^{ \beta} \gamma^{\mu} 
                  \gamma^{\nu} 
                  \gamma^{\rho} P_R c^{\beta})  + (-4 - 4 \e) Q_2^{\rm VLR},
\nnb \\ 
E_1^{\rm SLR} &=& 
(\bar{s}^{\alpha} \sigma_{\mu\nu} P_L d^{ \beta})
(\bar{u}^{ \beta} \sigma^{\mu\nu} P_R c^{\alpha}) - 6 \e Q_1^{\rm SLR},
\nnb \\ 
E_2^{\rm SLR} &=& 
(\bar{s}^{\alpha} \sigma_{\mu\nu} P_L d^{\alpha})
(\bar{u}^{ \beta} \sigma^{\mu\nu} P_R c^{\beta}) - 6 \e Q_2^{\rm SLR},
\nnb \\ 
E_1^{\rm SLL} &=& 
(\bar{s}^{\alpha} \gamma_{\mu} 
                  \gamma_{\nu} 
                  \gamma_{\rho} 
                  \gamma_{\sigma} P_L d^{\beta})
(\bar{u}^{ \beta} \gamma^{\mu} 
                  \gamma^{\nu} 
                  \gamma^{\rho} 
                  \gamma^{\sigma} P_L c^{\alpha})
              + (-64+96\e) Q_1^{\rm SLL} + (-16 + 8 \e) Q_3^{\rm SLL},
\nnb \\ 
E_2^{\rm SLL} &=& 
(\bar{s}^{\alpha} \gamma_{\mu} 
                  \gamma_{\nu} 
                  \gamma_{\rho} 
                  \gamma_{\sigma} P_L d^{\alpha})
(\bar{u}^{ \beta} \gamma^{\mu} 
                  \gamma^{\nu} 
                  \gamma^{\rho} 
                  \gamma^{\sigma} P_L c^{ \beta})
              + (-64 + 96 \e) Q_2^{\rm SLL} + (-16 + 8 \e) Q_4^{\rm SLL}.
\nnb 
\eea
The remaining evanescent operators (for the VRR, VRL, SRL and SRR sectors)
are obtained by interchanging $L$ and $R$ above.

\ \\
{\Large\bf Appendix C}
\def\theequation{C.\arabic{equation}}

This appendix contains definitions of the ``additional'' evanescent
operators that are {\em not} necessary as one-loop counterterms in the
$\Delta F=1$ effective Lagrangian in section \ref{sec:df1}. However,
they have to be included before performing transformation to the
``plus--minus'' basis in section \ref{sec:transf}.
\bea
E_3^{\rm VLL} &=& 
(\bar{s}^{\alpha} \gamma_{\mu} P_L c^{\alpha})
(\bar{u}^{ \beta} \gamma^{\mu} P_L d^{ \beta}) - Q_1^{\rm VLL},
\nnb\\
E_4^{\rm VLL} &=& 
(\bar{s}^{\alpha} \gamma_{\mu} P_L c^{ \beta})
(\bar{u}^{ \beta} \gamma^{\mu} P_L d^{\alpha}) - Q_2^{\rm VLL},
\nnb\\
E_5^{\rm VLL} &=& 
(\bar{s}^{\alpha} \gamma_{\mu} 
                  \gamma_{\nu} 
                  \gamma_{\rho} P_L c^{\alpha})
(\bar{u}^{ \beta} \gamma^{\mu} 
                  \gamma^{\nu} 
                  \gamma^{\rho} P_L d^{ \beta}) + (-16 + 4 \e) Q_1^{\rm VLL},
\nnb \\ 
E_6^{\rm VLL} &=& 
(\bar{s}^{\alpha} \gamma_{\mu} 
                  \gamma_{\nu} 
                  \gamma_{\rho} P_L c^{\beta})
(\bar{u}^{ \beta} \gamma^{\mu} 
                  \gamma^{\nu} 
                  \gamma^{\rho} P_L d^{\alpha}) + (-16 + 4 \e) Q_2^{\rm VLL},
\nnb \\[4mm]
E_3^{\rm VLR} &=& 
(\bar{s}^{\alpha} P_R c^{\alpha}) 
(\bar{u}^{ \beta} P_L d^{ \beta}) + \f{1}{2} Q_1^{\rm VLR},
\nnb\\
E_4^{\rm VLR} &=& 
(\bar{s}^{\alpha} P_R c^{ \beta})
(\bar{u}^{ \beta} P_L d^{\alpha}) + \f{1}{2} Q_2^{\rm VLR},
\nnb\\
E_5^{\rm VLR} &=& 
(\bar{s}^{\alpha} \sigma_{\mu\nu} P_R c^{\alpha})
(\bar{u}^{ \beta} \sigma^{\mu\nu} P_L d^{\beta})  + 3 \e Q_1^{\rm VLR},
\nnb \\ 
E_6^{\rm VLR} &=& 
(\bar{s}^{\alpha} \sigma_{\mu\nu} P_R c^{ \beta})
(\bar{u}^{ \beta} \sigma^{\mu\nu} P_L d^{\alpha}) + 3 \e Q_2^{\rm VLR},
\nnb \\[4mm]
E_3^{\rm SLR} &=& 
(\bar{s}^{\alpha} \gamma_{\mu} P_R c^{\alpha})
(\bar{u}^{ \beta} \gamma^{\mu} P_L d^{ \beta}) + 2 Q_1^{\rm SLR},
\nnb\\
E_4^{\rm SLR} &=& 
(\bar{s}^{\alpha} \gamma_{\mu} P_R c^{ \beta})
(\bar{u}^{ \beta} \gamma^{\mu} P_L d^{\alpha}) + 2 Q_2^{\rm SLR},
\nnb\\
E_5^{\rm SLR} &=& 
(\bar{s}^{\alpha} \gamma_{\mu} 
                  \gamma_{\nu} 
                  \gamma_{\rho} P_R c^{\alpha})
(\bar{u}^{ \beta} \gamma^{\mu} 
                  \gamma^{\nu} 
                  \gamma^{\rho} P_L d^{\beta})  + (8 + 8 \e) Q_1^{\rm SLR},
\nnb\\
E_6^{\rm SLR} &=& 
(\bar{s}^{\alpha} \gamma_{\mu} 
                  \gamma_{\nu} 
                  \gamma_{\rho} P_R c^{\beta})
(\bar{u}^{ \beta} \gamma^{\mu} 
                  \gamma^{\nu} 
                  \gamma^{\rho} P_L d^{\alpha}) + (8 + 8 \e) Q_2^{\rm SLR},
\nnb \\[4mm]
E_3^{\rm SLL} &=& (\bar{s}^{\alpha} P_L c^{\alpha})
              (\bar{u}^{ \beta} P_L d^{ \beta}) + \f{1}{2} Q_1^{\rm SLL}
                                                - \f{1}{8} Q_3^{\rm SLL},
\nnb\\
E_4^{\rm SLL} &=& (\bar{s}^{\alpha} P_L c^{ \beta})
              (\bar{u}^{ \beta} P_L d^{\alpha}) + \f{1}{2} Q_2^{\rm SLL}
                                                - \f{1}{8} Q_4^{\rm SLL},
\nnb\\
E_5^{\rm SLL} &=& (\bar{s}^{\alpha} \sigma_{\mu\nu} P_L c^{\alpha})
              (\bar{u}^{ \beta} \sigma^{\mu\nu} P_L d^{ \beta})        - 6 Q_1^{\rm SLL}
                                                                - \f{1}{2} Q_3^{\rm SLL},
\nnb\\
E_6^{\rm SLL} &=& (\bar{s}^{\alpha} \sigma_{\mu\nu} P_L c^{ \beta})
              (\bar{u}^{ \beta} \sigma^{\mu\nu} P_L d^{\alpha})        - 6 Q_2^{\rm SLL}
                                                                - \f{1}{2} Q_4^{\rm SLL},
\nnb \\ 
E_7^{\rm SLL} &=& 
(\bar{s}^{\alpha} \gamma_{\mu} 
                  \gamma_{\nu} 
                  \gamma_{\rho} 
                  \gamma_{\sigma} P_L c^{\alpha})
(\bar{u}^{ \beta} \gamma^{\mu} 
                  \gamma^{\nu} 
                  \gamma^{\rho} 
                  \gamma^{\sigma} P_L d^{ \beta})
               - 64 Q_1^{\rm SLL} + (-16 + 16 \e) Q_3^{\rm SLL},
\nnb \\
E_8^{\rm SLL} &=& 
(\bar{s}^{\alpha} \gamma_{\mu} 
                  \gamma_{\nu} 
                  \gamma_{\rho} 
                  \gamma_{\sigma} P_L c^{\beta})
(\bar{u}^{ \beta} \gamma^{\mu} 
                  \gamma^{\nu} 
                  \gamma^{\rho} 
                  \gamma^{\sigma} P_L d^{\alpha})
              - 64 Q_2^{\rm SLL} + (-16 + 16 \e) Q_4^{\rm SLL}.
\nnb 
\eea
The remaining evanescent operators (for the VRR, VRL, SRL and SRR sectors)
are obtained by interchanging $L$ and $R$ above.

\ \\
{\Large\bf Appendix D}
\def\theequation{D.\arabic{equation}}

In the present appendix, the notion of ``Greek projections''
\cite{B98,BW90,TV82} is recalled and generalized to the case of
SLL-sector operators. Let us denote the Dirac structure of the
operator in eq.~(\ref{generic.four.quark}) by ~$\Gamma_A \otimes
\Gamma_B$.~ The insertion of this operator in one- and two-loop
diagrams results in new Dirac structures like
\be
\Gamma_n \Gamma_A\,\otimes\, \Gamma^n \Gamma_B,
\ee
where $\Gamma_n = \gamma_{\mu_1} \gamma_{\mu_2} ... \gamma_{\mu_n}$.
Several examples of such structures occur in appendices A--C.  It
has been suggested in ref.~\cite{TV82} to project them onto physical
operators as follows. One defines the projection $G$ so that the
following equality is satisfied
\be
G\left[ \Gamma_n \Gamma_A\,\otimes\,\Gamma^n \Gamma_B \right] =
\xi G \left[ \,\Gamma^A\,\otimes\Gamma^B \right].\label{project1} 
\ee
In the case of ~$\Gamma_A=\Gamma_B=\gamma_\alpha P_L$,~ performing the
projection $G$ amounts to replacing $\otimes$ by $\gamma_\tau$ on both
sides of the above equation and contracting the indices using
$D$-dimensional Dirac algebra. In this manner, the coefficient $\xi$
is determined. One finds for instance:
\be
G \left[ 
(\bar s^\alpha \gamma_\mu \gamma_\nu \gamma_\rho P_L d^\beta )
(\bar u^\beta  \gamma^\mu \gamma^\nu \gamma^\rho P_L c^\alpha) \right] =
(16-4\e)~ G\left[ Q_1^{\rm VLL} \right] + {\cal O}(\e^2)
\label{greekproject}
\ee
with $Q_1^{\rm VLL}$ as defined in eq.~(\ref{normal.df1}). 

It has been pointed out in ref.~\cite{BW90} that for a proper
treatment of counterterms in two-loop calculations, one has to use
eq.~(\ref{greekproject}) only as a prescription for defining an
evanescent operator. In the case at hand, this is the operator
$E_1^{\rm VLL}$ of appendix B. As discussed in ref.~\cite{B98}, in the
case of VLR and SLR operators, the analogous projections are performed
by replacing $\otimes$ by 1 and $\gamma_\tau$, respectively.
Examples of the corresponding evanescent operators can be found in
appendices A--C.

The projections in the SLL sector are slightly more involved.  In the
case of the insertion of $Q_1^{\rm SLL}$ or $Q_3^{\rm SLL}$, the
r.h.s. of eq.~(\ref{project1}) has to be generalized to a linear
combination of these two operators.  The same applies to the pair
$(Q_2^{\rm SLL},\,Q_4^{\rm SLL})$. The projection $G$ is now performed
by replacing $\otimes$ ~by~ $\gamma_\alpha \gamma_\beta$. After the
projection, one finds linear combinations of $g_{\alpha \beta}$ and
$\gamma_\alpha \gamma_\beta$ on both sides of the equation. This
allows extracting the coefficients in question.  One finds for
instance
\mathindent0cm
\be
G \left[ 
(\bar s^\alpha \gamma_\mu \gamma_\nu \gamma_\rho \gamma_\sigma P_L d^\beta )
(\bar u^\beta  \gamma^\mu \gamma^\nu \gamma^\rho \gamma^\sigma P_L c^\alpha) 
\right] =   (64-96 \e)~ G\left[ Q_1^{\rm SLL}\right] 
          + (16-8 \e)~ G\left[ Q_3^{\rm SLL} \right] + {\cal O}(\e^2).~~~
\ee
\mathindent1cm
The corresponding evanescent operator is $E_1^{\rm SLL}$ in appendix B.
An alternative approach to projections can be found in
ref.~\cite{HN95}.

\newpage
{\Large\bf Appendix E}
\def\theequation{E.\arabic{equation}}

In this appendix, the $1/\e$ and $1/\e^2$ poles in the one- and
two-loop diagrams are given for the $\Delta F=1$ calculation in the
SLL sector.  Analogous results for the remaining sectors can be found
in refs.~\cite{BW90} and \cite{BJLW93}. The gauge-fixing parameter
$\lambda$ is set to unity here, i.e. the Feynman--'t~Hooft gauge is
used.

Each insertion results in a linear combination of $Q_1^{\rm SLL}$, ...,
$Q_4^{\rm SLL}$, after subtracting the evanescent counterterms (see
appendix B) or, alternatively, after performing the ``Greek
projections'' (see appendix D). Table~\ref{table1} gives the
singularities (without colour factors) in the coefficients of the
resulting operators, for each diagram separately. The numbering of the
diagrams and values of the colour factors are exactly as in figs.~1,~2
and tables~1,~2 of ref.~\cite{BW90}. The multiplicity factors of the
diagrams are included.

In the two-loop case, the singularities include one-loop diagrams with
counterterm insertions. The counterterms proportional to evanescent
operators are multiplied by an additional factor $1/2$, and, at the
same time, the term ~$-2\hat{b}\hat{c}$~ in eq.~(\ref{gamma1}) is
ignored. Correctness of such a trick has been justified in
refs.~\cite{BW90,DG91}.

The singularities from table~\ref{table1} apply for the pair
$(Q_2^{\rm SLL},\,Q_4^{\rm SLL})$, too. After including colour factors and
summing the diagrams, the $1/\e$ singularities build a $4\times 4$
matrix in the basis $\{Q_1^{\rm SLL},\,Q_2^{\rm SLL},\,Q_3^{\rm SLL},\,Q_4^{\rm SLL}\}$
\be
\hat B = \frac{\alpha_s}{4\,\pi}\,\hat B_1 + 
         \left(\frac{\alpha_s}{4\,\pi}\right)^2\,\hat B_2+{\cal O}(\alpha_s^3),
\ee
from which the anomalous-dimension matrix can be
obtained by means of
\bea \label{gB0}
\gamma^{(0)\rm SLL}_{ij} &=& -2 \left[2\,a_1\,\delta_{ij}+
                                    (\hat B_1)_{ij}\right],\\
\label{gB1}
\gamma^{(1)\rm SLL}_{ij} &=& -4 \left[2\,a_2\,\delta_{ij}+
                                    (\hat B_2)_{ij}\right].
\eea
Here, $a_1$ and $a_2$ originate from $1/\e$ singularities in the 
quark field renormalization constants. They read
\be
a_1=-C_F,\qquad a_2=C_F\left[\frac34\,C_F-\frac{17}{4}\,N+\frac12\,f\right].
\ee
Remembering the trick applied to evanescent operators here, it is easy
to verify that eqs.~(\ref{gB0}) and (\ref{gB1}) are equivalent to
eqs.~(\ref{gamma0}) and (\ref{gamma1}) from section \ref{sec:df2}.

\begin{table}[ht]
\begin{center}
{\footnotesize
\begin{tabular}{|c|c||c|c|c|c||c|c|c|c|}
\hline
\TT     
    &   &\mc{4}{c||}{$Q_1^{\rm SLL}$}&\mc{4}{|c|}{$Q_3^{\rm SLL}$}\\
\hline
\TT 
$D$ &$M$&\mc{2}{c|}{$Q_1^{\rm SLL}$ and $Q_2^{\rm SLL}$}
        &\mc{2}{c||}{$Q_3^{\rm SLL}$ and $Q_4^{\rm SLL}$}
        &\mc{2}{|c}{$Q_1^{\rm SLL}$ and $Q_2^{\rm SLL}$}
        &\mc{2}{|c|}{$Q_3^{\rm SLL}$ and $Q_4^{\rm SLL}$}\\
\hline
\TT  
    &  &$1/\e^2$  &$1/\e$  &$1/\e^2$  &$1/\e$  &$1/\e^2$  &$1/\e$  &$1/\e^2$&$1/\e$\\
\hline
\hline
\TT 
$ 1$&$2 $&$-    $&$8    $&$-     $&$0      $&$-      $&$0      $&$-     $&$0   $\\
$ 2$&$2 $&$-    $&$-2   $&$-     $&$-1/2   $&$-      $&$-24    $&$-     $&$-6  $\\
$ 3$&$2 $&$-    $&$2    $&$-     $&$-1/2   $&$-      $&$-24    $&$-     $&$6   $\\
\hline
\TT
$ 4$&$2 $&$-16  $&$16   $&$0     $&$0      $&$0      $&$0      $&$0     $&$0   $\\
$ 5$&$2 $&$-4   $&$9    $&$-1    $&$5/4    $&$-48    $&$76     $&$-12   $&$7   $\\
$ 6$&$2 $&$-4   $&$9    $&$1     $&$-7/4   $&$48     $&$-52    $&$-12   $&$7   $\\
$ 7$&$2 $&$0    $&$-4   $&$0     $&$0      $&$0      $&$0      $&$0     $&$4   $\\
$ 8$&$2 $&$0    $&$2    $&$0     $&$-1/2   $&$0      $&$-24    $&$0     $&$-2  $\\
$ 9$&$2 $&$0    $&$2    $&$0     $&$1/2    $&$0      $&$24     $&$0     $&$-2  $\\
$10$&$4 $&$-8   $&$-8   $&$0     $&$0      $&$0      $&$0      $&$0     $&$4   $\\
$11$&$4 $&$2    $&$0    $&$1/2   $&$5/4    $&$24     $&$20     $&$6     $&$8   $\\
$12$&$4 $&$-2   $&$0    $&$1/2   $&$5/4    $&$24     $&$20     $&$-6    $&$-4  $\\
$13$&$4 $&$8    $&$-4   $&$0     $&$0      $&$0      $&$0      $&$0     $&$0   $\\
$14$&$4 $&$-2   $&$0    $&$-1/2  $&$-1/4   $&$-24    $&$28     $&$-6    $&$0   $\\
$15$&$4 $&$2    $&$0    $&$-1/2  $&$-1/4   $&$-24    $&$28     $&$6     $&$-4  $\\
$16$&$4 $&$8    $&$-4   $&$0     $&$0      $&$96     $&$64     $&$0     $&$0   $\\
$17$&$4 $&$8    $&$4    $&$2     $&$2      $&$0      $&$0      $&$0     $&$0   $\\
$18$&$4 $&$-8   $&$4    $&$0     $&$0      $&$96     $&$64     $&$0     $&$0   $\\
$19$&$4 $&$-8   $&$-4   $&$2     $&$2      $&$0      $&$0      $&$0     $&$0   $\\
$20$&$4 $&$-4   $&$10   $&$-1    $&$1      $&$48     $&$-64    $&$12    $&$2   $\\
$21$&$4 $&$-4   $&$10   $&$1     $&$-2     $&$-48    $&$112    $&$12    $&$-22 $\\
$22$&$1 $&$-16  $&$0    $&$0     $&$0      $&$0      $&$0      $&$0     $&$0   $\\
$23$&$1 $&$-4   $&$5    $&$-1    $&$1/4    $&$-48    $&$28     $&$-12   $&$-5  $\\
$24$&$1 $&$-4   $&$5    $&$1     $&$-3/4   $&$48     $&$-4     $&$-12   $&$-5  $\\
$25$&$4 $&$24   $&$-20  $&$0     $&$0      $&$0      $&$0      $&$0     $&$0   $\\
$26$&$4 $&$-6   $&$2    $&$-3/2  $&$-1/4   $&$-72    $&$108    $&$-18   $&$6   $\\
$27$&$4 $&$6    $&$-2   $&$-3/2  $&$-1/4   $&$-72    $&$108    $&$18    $&$-18 $\\
$28$&$4 $&$0    $&$0    $&$0     $&$3      $&$0      $&$-144   $&$0     $&$0   $\\
$29$&$2 $&$\!\!${\scriptsize{$-5 N+2 f$}}&$\!\!\f{26 N}3-\f{8 f}3$&$0$&$0  $&$0$&     
  $0           $&$\!\! \f{5 N}3-\f{2 f}3$&$\!\! -\f{16 N}9+\f{4 f}9$\\
$30$&$2 $&$0           $&$0         $&$\!\! \f{5 N}{12}-\f{f}{6}$&$\!\! -\f{17 N}{72}+\f{f}{36}$&
  $\!\!${\scriptsize{$20 N - 8 f$}}&$\!\! -\f{134 N}3+\f{44 f}3$&$\!\! \f{10 N}3-\f{4 f}3$&
  $\!\! -\f{32 N}9+\f{8 f}9$\\
\TT
$31$&$2 $&$0           $&$0         $&$\!\! \f{5 N}{12}-\f{f}{6}$&$\!\! -\f{17 N}{72}+\f{f}{36}$&
  $\!\!${\scriptsize{$20 N - 8 f$}}&$\!\! -\f{134 N}3+\f{44 f}3$&$\!\! -\f{10 N}3+\f{4 f}3$&
  $\!\! \f{62 N}9-\f{20 f}9$\\
\hline
\end{tabular}
}
\caption{\footnotesize
\label{table1} Pole parts of the one- and two-loop diagrams
with insertions of $Q_1^{\rm SLL}$ and $Q_3^{\rm SLL}$.  The colour factors
are omitted, whereas the multiplicity ($M$) is taken into account. The
numbering is according to fig. 2 of ref. \cite{BW90}.  While the
singularities in front of the resulting $Q_1^{\rm SLL}$ and $Q_2^{\rm SLL}$
are the same in this table, they become different after the inclusion
of colour factors. The same comment applies to $Q_3^{\rm SLL}$ and
$Q_4^{\rm SLL}$. When the colour factors are omitted, the results for
$Q_2^{\rm SLL}$ and $Q_4^{\rm SLL}$ insertions are equal to those for
$Q_1^{\rm SLL}$ and $Q_3^{\rm SLL}$ insertions, respectively.}
\end{center}
\end{table}

\clearpage
\newpage 

{\Large\bf Appendix F}
\def\theequation{F.\arabic{equation}}

In the present appendix, we outline details of our determination of
the penguin-diagram generated ADMs in eqs.~(\ref{ga0peng}) and
(\ref{ga1peng}). Let us begin with the SM QCD-penguin operators $Q_3$,
\ldots $Q_6$ that are listed in eq.~(\ref{SM.ops}). Their $4\times 4$
ADM can be split into contributions from the current-current and
penguin diagrams: $\hat{\gamma}^{\rm SM} = \hat{\gamma}^{\rm SM}_{cc} +
\hat{\gamma}^{\rm SM}_p$. The structure of $Q_3$, \ldots $Q_6$ implies that
\be \label{gammaSMcc}
\hat{\gamma}^{\rm SM}_{cc} = \left( \begin{array}{cccc}
\gamma^{\rm VLL}_{22} & \gamma^{\rm VLL}_{21} & 0 & 0 \\[1mm] 
\gamma^{\rm VLL}_{12} & \gamma^{\rm VLL}_{11} & 0 & 0 \\[1mm]
0 & 0 & \gamma^{\rm VLR}_{22} & \gamma^{\rm VLR}_{21} \\[1mm]
0 & 0 & \gamma^{\rm VLR}_{12} & \gamma^{\rm VLR}_{11}
\end{array} \right),
\ee
where $\hat{\gamma}^{\rm VLL}$ and $\hat{\gamma}^{\rm VLR}$ up to the two-loop
level are given in eqs.~(3.2)--(3.5). The matrix
$\hat{\gamma}^{\rm SM}_p$ is most easily found by subtracting
$\hat{\gamma}^{\rm SM}_{cc}$ from the full $\hat{\gamma}^{\rm SM}$. One- and
two-loop contributions to the latter matrix in the NDR--$\overline{\rm
MS}$ are listed in table 5 of ref.~\cite{BJLW93}.\footnote{
They coincide with results of independent calculations in refs.~\cite{CFMR94} and \cite{CMM98.df}.}
This way one finds
\bea
\hat{\gamma}^{{\rm SM}\,(0)}_p &=&
\begin{array}{c} \left(\f{4}{3}, \f{2f}{3}, 0, \f{2f}{3} \right)^T \times
\left(-\f{1}{N},1,-\f{1}{N},1 \right), \end{array} \nnb\\[2mm]
\hat{\gamma}^{{\rm SM}\,(1)}_p &=& \left( \begin{array}{ccccccc}
-\f{64}{27} + \f{172}{27 N^2} && 
-\f{460}{27 N} + \f{352 N}{27} &&
-\f{244}{27} - \f{188}{27 N^2} &&
\f{260}{27 N} + \f{172 N}{27} \\[1mm]
-\f{4}{3 N} + 6 N &&
-\f{14}{3} &&
\f{32}{3 N} - 6 N && 
-\f{14}{3}\\[1mm]
0 && 0 && 0 && 0 \\[1mm]
0 && 0 && 0 && 0 \\[1mm]
\end{array} \right)\nnb \\[2mm]
&+& \left( \begin{array}{ccccccc}
-\f{8}{3 N} + 3 N && 
-\f{1}{3} &&
\f{10}{3 N} - 3 N &&
-\f{1}{3} \\[1mm]
-\f{20}{27} + \f{74}{27 N^2} &&
-\f{164}{27 N} + \f{110 N}{27} &&
-\f{56}{27} + \f{2}{27 N^2} && 
-\f{20}{27 N} + \f{74 N}{27} \\[1mm]
\f{20}{3 N} - 3 N&&
-\f{11}{3} &&
\f{2}{3 N} + 3 N &&
-\f{11}{3} \\[1mm]
-\f{56}{27} - \f{178}{27 N^2} && 
\f{250}{27 N} - \f{16 N}{27} &&
\f{70}{27} + \f{74}{27 N^2} && 
-\f{254}{27 N} + \f{110 N}{27} 
\end{array} \right) f. \label{gammaSMp}
\eea

Let us now extend the considered operator set to $\{ Q_3, Q_4, Q_5,
Q_6, Q_{11}, Q'_{11}, Q_{13}, Q_{12} \}$, where the extra operators
have been defined in eqs.~(\ref{additional}) and (\ref{addprime}). At
this point, we ignore the fact that $Q_{11}$ and $Q'_{11}$ are related
by a Fierz relation in $D=4$, i.e.\ we treat both of them as
independent normal (non-evanescent) operators.

The four extra operators $\{ Q_{11}, Q'_{11}, Q_{13}, Q_{12} \}$ can
be obtained from $\{ Q_3, Q_4, Q_5, Q_6 \}$ by skipping the
$u$-, $c$- and $b$-quarks in the sum over flavours. Consequently, the full
$8\times 8$ ADM up to two loops takes the form
\be 
\hat{\gamma}^{\ }_{8 \times 8} = \left( \begin{array}{cc}
\hat{\gamma}^{\rm SM}         & 0_{4 \times 4} \\[2mm] 
\hat{\gamma}^{\rm SM}_{p,f=2} & \hat{\gamma}^{\rm SM}_{cc} 
\end{array} \right).
\ee
We have tacitly assumed here that each of our operators is accompanied
by a corresponding one-loop evanescent operator containing triple
products of Dirac matrices inside the quark currents, defined in full
analogy to $E_1^{\rm VLL}$, $E_2^{\rm VLL}$, $E_1^{\rm VLR}$ and $E_2^{\rm VLR}$ in
appendix B, including the coefficients at the ${\mathcal O}(\e)$ terms
there.

In the next step, we transform our $8 \times 8$ ADM to the basis
$\{ Q_3, Q_4, Q_5, Q_6, Q_{11}, Q_{12}, Q_{13}, E_{11} \}$, where $E_{11}
= Q'_{11}-Q_{11}$ is still treated as a normal (non-evanescent)
operator. The transformed ADM reads
\mathindent0cm
\be
\hat{\gamma}'_{8 \times 8} = \hat{R}\, \hat{\gamma}^{\ }_{8 \times 8}\, \hat{R}^{-1}
\hspace{5mm} \mbox{with} \hspace{5mm}
\hat{R} = \left( \begin{array}{cc}
1_{4 \times 4} & 0_{4 \times 4} \\[2mm] 
0_{4 \times 4} & X_{4 \times 4} 
\end{array} \right)
\hspace{5mm} \mbox{and} \hspace{5mm}
X_{4 \times 4} = \left( \begin{array}{cccc}
 1 &  0 & 0 & 0\\
 0 &  0 & 0 & 1\\
 0 &  0 & 1 & 0\\
-1 & 1 & 0 & 0
\end{array} \right).
\ee
\mathindent1cm

Finally, we depart from the ${\rm MS}$ scheme for $E_{11}$ (still
thinking of it as of a normal operator though) by introducing finite
terms in the one-loop renormalization constants that correspond to its
mixing via penguin diagrams into $Q_3$, \ldots $Q_6$. This amounts to
replacing $a^{11}_{ik}$ in eq.~(\ref{counterterms}) by $a^{11}_{ik} +
\e a^{01}_{ik}$ when the index $i$ corresponds to $E_{11}$, and the
index $k$ corresponds to $Q_3$, \ldots $Q_6$. We adjust
\be \label{a01}
a^{01}_{E_{11},Q_3} = a^{01}_{E_{11},Q_5} = -\f{2}{3N}
\hspace{1cm} \mbox{and} \hspace{1cm}
a^{01}_{E_{11},Q_4} = a^{01}_{E_{11},Q_6} = \f{2}{3}
\ee
to make the renormalized one-loop penguin matrix element of $E_{11}$
vanish. The one-loop ADM remains intact, while the resulting
transformation of the two-loop ADM reads (c.f. eq.~(\ref{transf1}))
\mathindent0cm
\be \label{last.transf}
\hat{\gamma}''^{(1)}_{8 \times 8} = \hat{\gamma}'{}^{(1)}_{8 \times 8}
+ \left[ \Delta \hat{r}, \hat{\gamma}'^{(0)}_{8 \times 8} \right]
        + 2 \beta_0 \Delta \hat{r}
\hspace{5mm} \mbox{with} \hspace{5mm}	
(\Delta r)_{ij} = \delta_{i8} \left[ -\f{2}{3N} \left( \delta_{j1} + \delta_{j3} \right)
                                     +\f{2}{3}  \left( \delta_{j2} + \delta_{j4} \right) \right]\!.~
\ee
\mathindent1cm
At this point, our non-${\rm MS}$ scheme with $E_{11}$ treated as a
normal operator becomes equivalent to the ${\rm MS}$ scheme with
$E_{11}$ treated as an evanescent operator.\footnote{
Strictly speaking, this is true only up to finite renormalization in
the one-loop mixing of evanescent operators among themselves, which
should be absent in the ${\rm MS}$ scheme. However, such a finite
renormalization has no effect on the ADM of the physical operators up
to two loops.}
We explicitly verify that the 8th row of $\hat{\gamma}''_{8 \times 8}$
up to two loops has only a single non-vanishing entry that corresponds
to the mixing of $E_{11}$ with itself. Consequently, the Wilson
coefficient of $E_{11}$ has no effect on the RG evolution of the
Wilson coefficients of normal operators, as it should be for any
evanescent operator in the ${\rm MS}$ scheme. Let us note that it would
not be the case if the transformation (\ref{last.transf}) was not performed.

To find the actual ADM for the normal operators only (now with
$E_{11}$ treated as an evanescent operator, and in the ${\rm MS}$
scheme), we remove the 8th row and 8th column from the matrix
$\hat{\gamma}''_{8 \times 8}$. The resulting $7 \times 7$ matrix
reads
\be \label{gamma77}
\hat{\gamma}^{\ }_{7 \times 7} = \left( \begin{array}{cc}
\hat{\gamma}^{\rm SM} & 0_{4 \times 3} \\[2mm] 
\hat{\gamma}_p    & \hat{\gamma}_{cc} 
\end{array} \right),
\ee
where the one- and two-loop contributions to $\hat{\gamma}^{\rm SM}$,
$\hat{\gamma}_{cc}$ and $\hat{\gamma}_p$ coincide with what has been
already given in eqs.~(\ref{gammaSMcc})-(\ref{gammaSMp}),
(\ref{gammacc}) and (\ref{ga0peng})-(\ref{ga1peng}), respectively. Our
final results for $\hat{\gamma}_p$ in
eqs.~(\ref{ga0peng})-(\ref{ga1peng}) have actually been extracted from
eq.~(\ref{gamma77}). As far as $\hat{\gamma}_{cc}$ in
eq.~(\ref{gammacc}) is concerned, eq.~(\ref{gamma77}) gives us a nice
confirmation of the result that has in practice been determined using
a much simpler method.

The transformation (\ref{last.transf}) was missed in the original
version of our paper in 2000. The mistake was pointed out in
ref.~\cite{Morell:2024aml}. The current appendix was added in 2024,
simultaneously with implementing the proper correction in
eq.~(\ref{ga1peng}).

The reader might wonder why our finite subtraction in
eq.~(\ref{a01}) was restricted to the penguin matrix elements
only, i.e.\ why no similar operation was necessary for the one-loop
current-current matrix element of $E_{11}$. It was the case because
such a matrix element turns to vanish after subtracting the evanescent
counterterms only, and passing to $D=4$, as already discussed in
section~\ref{sec:transf}, below eq.~(\ref{opm.VLL.LR}).

\setlength {\baselineskip}{0.2in}  
\end{document}